\RequirePackage{fix-cm}
\documentclass[smallextended]{svjour3}       
\smartqed  
\usepackage{graphicx}
\usepackage{multirow}%
\usepackage{amsmath,amssymb,amsfonts}%
\usepackage{mathrsfs}%
\usepackage[title]{appendix}%
\usepackage{xcolor}%
\usepackage{textcomp}%
\usepackage{manyfoot}%
\usepackage{booktabs}%
\usepackage{algorithm}%
\usepackage{algorithmicx}%
\usepackage{algpseudocode}%
\usepackage{listings}%
\usepackage{newtxmath}
\usepackage{etoolbox}
\usepackage{bm}
\usepackage{xcolor}
\usepackage{subfigure}
\usepackage{todonotes}
\usepackage{float}
\usepackage{ulem}
\usepackage{xcolor}
\usepackage{soul}

\usepackage[round,sort&compress]{natbib}
\definecolor{highlightcolor}{RGB}{255,255,255}
\begin{document}

\title{Modeling gas flows in packed beds with the lattice Boltzmann method: validation against experiments}
\titlerunning{LBM simulation and experiments of gas flow in packed bed}

\author{Tanya Neeraj*\textsuperscript{1} \thanks{Corresponding author: Tanya Neeraj (tanya.neeraj@ovgu.de)}\and
Christin Velten\textsuperscript{1} \and
Gabor Janiga\textsuperscript{1} \and
Katharina Z\"ahringer\textsuperscript{1} \and
Reza Namdar\textsuperscript{3} \and
Fathollah Varnik\textsuperscript{3} \and
Dominique Th\'evenin\textsuperscript{1} \and
Seyed Ali Hosseini\textsuperscript{1,2}}
\institute{\textsuperscript{1}Laboratory of Fluid Dynamics and Technical Flows, University of Magdeburg ``Otto von Guericke'', Magdeburg, 39106, Germany.\\
\textsuperscript{2}Department of Mechanical and Process Engineering, ETH Z\"urich, 8092, Z\"urich, Switzerland.\\
\textsuperscript{3} Interdisciplinary Centre for Advanced Materials Simulation (ICAMS), Ruhr-University Bochum, Bochum, 44801, Germany.
}
\date{Received: date / Accepted: date}
\maketitle
\begin{abstract}
This study aims to validate the lattice Boltzmann method and assess its ability to accurately describe the behavior of gaseous flows in packed beds. To that end, simulations of a model packed bed reactor, corresponding to an experimental bench, are conducted, and the results are directly compared with experimental data obtained by Particle Image Velocimetry measurements. It is found that the lattice Boltzmann solver exhibits very good agreement with experimental measurements. Then, the numerical solver is further used to analyze the effect of the number of packing layers on the flow structure and to determine the minimum bed height above which the changes in flow structure become insignificant. Finally, flow fluctuations in time are discussed. The findings of this study provide valuable insights into the behavior of the gas flow in packed bed reactors, opening the door for further investigations involving additionally chemical reactions, as found in many practical applications.
\end{abstract}

\section{Introduction}
\label{intro}
Packed bed reactors are an essential component of many engineering, chemical and petrochemical processes. Their relevance lies in their ability to provide a large surface area for mass transfer, high pressure drop, and high heat transfer rates. Investigating these packed beds provides valuable insights into their performance and efficiency, crucial for optimizing their design and operation. The flow behavior, pressure drop, heat transfer, and mass transfer characteristics of packed beds are being analyzed and this information then aids in predicting their performance under different operating conditions and facilitates the development of more efficient and cost-effective industrial processes. These reactors consist of a fixed bed involving many solid particles, often reacting, in contact with the surrounding liquid or gas phase. In the present study, only gas flows will be considered, since they correspond to the high-temperature applications (drying, heat storage, roasting, calcination, pyrolysis\ldots) driving the present investigations. Packed bed reactors allow for efficient heat and mass transfer and -- as a consequence -- reactions between gas and solid phase. The combustion and reaction processes in these packed bed reactors are intricate and involve a number of coupled physical and chemical phenomena -- most prominently mass transfer, heat transfer, and fluid dynamics~\citep{Bao1,Dixon,Young,Aglave}.\\

Unfortunately, the full experimental characterization of these packed bed geometries faces several challenges including the complex and three-dimensional nature of the beds, tracking individual particles for moving beds, capturing non-uniformities and heterogeneities of the flow within the bed, scaling up experimental findings, and ensuring reproducibility. Additionally, high-temperature processes in packed beds cannot be investigated experimentally with most existing diagnostics, either because of the very difficult access to the inter-particle spaces and/or of the harsh conditions found there. Getting accurate experimental data regarding the exchange processes at this level is essential to optimize further all underlying applications, improving energy efficiency and reducing environmental impact~\citep{Bao1,ALOBAID2022100930}.\\

As a complementary source of information, sophisticated mathematical and computational models should be used to accurately simulate the behavior of packed bed reactors. With the advent of computational fluid dynamics (CFD) and constant progress in computational power, there is a growing interest in modeling numerically all kinds of multiphase flows~\citep{bpg,Kun1,Kun2}, supporting also design, operation, and optimization of packed bed reactors~\citep{Viktor,ALOBAID2022100930,freund2003numerical,Augier}. Numerical methods, such as the lattice Boltzmann method (LBM), can help in the experimental characterization of these packed bed geometries by providing a complementary approach to experimental techniques. LBM is an approach that is rapidly gaining popularity due to its ability to simulate a wide variety of complex flows. It is extremely well-suited for describing fluid flows in porous media~\citep{Bao1,Chen,amir16}. It simulates the movement of particles on a discrete lattice ~\citep{sudhakar2020evolution,Yang} and can provide information about the flow behavior, velocity distribution, and mass transfer characteristics within the packed bed at the pore-scale level. LBM simulations excel in capturing the complexities of three-dimensional packed bed structures and provide a means to visualize internal flow patterns that are inaccessible experimentally. It can also be used to optimize the design and operation of packed beds by predicting their performance under different operating conditions. This method is particularly efficient for complex geometries due to a very easy grid generation and treatment of boundary conditions. Additionally, LBM simulations can be scaled up to larger bed sizes (as done in the present study), aiding in the prediction of packed bed performance in industrial settings. In summary, LBM is in principle very simple, shows excellent scalability on parallel supercomputers, and extensions have been proposed to represent accurately all kinds of flows~\citep{sudhakar2020evolution,Kruger,HE2019160,boivin_comb,boivin_compres}. Therefore, by integrating numerical methods like LBM with experimental data, we can gain a more comprehensive understanding of packed bed behavior, optimize their design, and enhance the efficiency of the industrial processes.\\

However, to fully leverage the benefits of simulations, it is imperative to ensure the accuracy and reliability of the results obtained. Even with sophisticated models and computational techniques, there is a need to establish trust in numerical outcomes. This necessitates the combination of experimental and numerical studies for benchmarking, as experimental validation serves as a critical reference for assessing the accuracy of numerical simulations. Here also, many challenges must be solved to ensure sufficiently accurate predictions for such complex two-phase flows while keeping computational requirements at an acceptable level. Due to non-linear coupling processes between momentum, heat, and mass transfer, very slight changes in geometry at small scale or minute modifications of the operating conditions may have a significant impact on the behavior of the whole reactor~\citep{Prokopova,HUSSAIN2015143}.\\

In recent decades, the lattice Boltzmann method (LBM) has proved to be an interesting numerical approach to address this type of phenomena. \cite{RONG2014508}, for example, used LBM to explore the effects of particle size distribution on flow behavior and drag forces within binary mixtures of particles. They identified flow characteristics related to pore geometries and proposed a new equation for calculating mean individual drag force. Another study by the same authors \cite{RONG2015146} used LBM to simulate fluid flow through packed beds of uniform ellipsoids and investigated the impact of particle shape, orientation, and volume fraction on flow behavior and pressure drop. \cite{Sullivan} on the other hand, explored the application of lattice Boltzmann methods for simulating hydrodynamics, reaction, and mass transfer in a packed bed reactor of catalyst particles at pore-scale. Spatially varying values of diffusivity have been considered to differentiate between intra- and inter-particle diffusion. A similar study used LBM to simulate concentration profiles in a fixed bed filled with spherical porous adsorbents~\citep{VERMA20073685}. The central objective was to investigate how concentration profiles change due to inter- and intra-particle mass transport in an adsorber with small tube-to-sphere diameter ratios, down to less than 10. Various packing arrangements and different pore diffusivities were considered, validated with experimental data obtained by near-infrared optical tomography in a tubular adsorber packed with zeolite particles. \cite{TARIQ2022107382} investigated flow structures at the level of the inter-particle space, without discussions of the overall flow features within and above a full bed of particles.

It is noteworthy that, despite the large variety of lattice Boltzmann studies of the flow through complex geometries including packed beds, studies, which also include an experimental validation of numerical results are quite sparse. Combined experimental-numerical investigations proposed in the present study thus aims at filling this gap.
Considering the complexity of this problem, combining experimental and numerical investigations of exactly the same configuration appears to be the best solution in order to push further our understanding. This is the central objective of the present study. It was already stressed that experimental measurements are extremely challenging due to limited access and high temperatures. Avoiding in this first investigation the latter issue by considering only a two-phase cold flow, a solution must still be found to measure accurately flow properties in-between particles within a packed bed.\\

Again, at the difference of most published studies, the present investigation concentrates only on {\em gaseous} flows through packed beds. To the best of our knowledge, no publication using LBM considered such gas flows in large-scale packed beds up to now. Therefore, and even if the in-house LBM solver employed in this study (ALBORZ, described in particular in~\cite{Hosseini,TGV,hosseini2021central}) has been previously validated for a variety of configurations, it is particularly important to quantify again its accuracy and reliability by direct comparisons with experimental data for the same packed-bed configuration. In what follows, the air flow within a body-centered-cubic (BCC) packing will be compared to results from PIV measurements regarding flow structures and velocities, both directly above as well as within the packed bed. This will deepen our understanding of the flow behavior inside the reactor in order to push further process design, ultimately enhancing performance and efficiency. Two different particle Reynolds numbers will be examined in a range relevant for practical applications.
In addition, the numerical results will be analyzed to better understand the overall process as well as the flow behavior at selected positions. The occurrence of flow fluctuations and instabilities will finally be discussed. The LBM results show good agreement with experimental data, providing a strong basis for the numerical model developed in this study.

\section{Numerical solver ALBORZ: the lattice Boltzmann method}
\subsection{Modified central Hermite multiple-relaxation time lattice Boltzmann solver}
The lattice Boltzmann method is a solver for the Boltzmann equation in the limit of the hydrodynamic regime~\citep{Xiaoyi}, tailored to recover the Navier-Stokes equations. It can be derived by first discretizing the Boltzmann equation in phase space, the space of particle degrees of freedom, via projection onto Hermite polynomials and the use of the Gauss-Hermite quadrature. The resulting system of coupled hyperbolic equations is then integrated along characteristics~\citep{hosseini2023lattice}. This in turn yields the now-famous "stream-collide" equation:
\begin{equation}
    f_i\left(\bm{r}+\bm{c}_i\delta t, t+\delta t\right) - f_i\left(\bm{r}, t\right) = \Omega_i,
\end{equation}
for discrete distribution functions $f_i$, where $\bm{c}_i$ are discrete particle velocities, $\Omega_i$ is the discrete collision operator, modeling equilibration of the distribution function due to molecular collisions, and $\delta t$ is the time-step. Close to local thermodynamic equilibrium (LTE) conditions (which is valid for small Knudsen numbers, as found in all flows considered here) a linear relaxation operator (Bhatnagar-Gross-Krook, BGK) is usually employed, which in its discrete form reads
\begin{equation}
    \Omega_i = \omega\left( f^{\rm eq}_i - f_i\right) + \Xi_i,
\end{equation}
where $\omega$ is the relaxation frequency coefficient and $f^{\rm eq}_i$ is the discrete form of the continuous attractor obtained via minimization of entropy, i.e., the Maxwell-Boltzmann distribution~\citep{chapman1990mathematical}. It must be noted that here the collision operator is supplemented with a term $\Xi_i$ accounting for deviations in the diagonal components of the third-order moments of the equilibrium term (for models relying on third-order quadratures). Details on this term can be found in~\cite{feng2015three,prasianakis2007lattice,hosseini2020compressibility}, among other sources. In the context of the lattice Boltzmann method the discrete equilibrium distribution function is expressed as a finite-order expansion of the Maxwell-Boltzmann distribution in terms of Hermite polynomials by means of the Gauss-Hermite quadrature:
\begin{equation}
    f^{\rm eq}_i = w_i \sum_{n=0}^{N}\frac{1}{n!c_s^{2n}} \mathcal{H}_n(\bm{c}_i):a_n^{\rm eq}(\rho,\bm{u},p/\rho),
\end{equation}
where $\mathcal{H}_n$ is the Hermite polynomial tensor of order/rank $n$, $a_n^{\rm eq}$ the corresponding equilibrium coefficient, $w_i$ are weights associated with the quadrature and $c_s$ the lattice speed of sound, also tied to the quadrature. Further quantities appearing in the system are $\rho$ the fluid density, $\bm{u}$ the velocity, $p$ the pressure and $N$ the order of expansion. For the remainder of the manuscript we will restrict ourselves to the D2Q9 and D3Q27 stencils, i.e. third-order quadratures, where $D$ is the physical dimension (2D or 3D), and Q the number of discrete velocities. In two dimensions, $N$ is set to four, while in 3D it is set to six. Both are dictated by the maximum number of terms supported by the stencils. Contrary to classical second-order polynomial expansions, the present model allows the discrete form to preserve Galilean invariance of the dissipation rate of shear modes at the Navier-Stokes level, which is an essential aspect. The term $\Xi_i$ restores Galilean invariance to the dissipation rate of normal modes and is defined as~\citep{hosseini2020compressibility,feng2015three}:
\begin{equation}\label{eq:correction}
    \Xi_i = \left(1-\frac{\omega}{2}\right)\frac{w_i}{2c_s^4}\left(\bm{\nabla}\mathcal{H}_2\right):\Delta^{\rm eq}_3,
\end{equation}
where $\Delta^{\rm eq}_3$ is a diagonal tensor of rank 3 defined as:
\begin{equation}
    \Delta^{\rm eq}_{\alpha\alpha\alpha} = \rho u_\alpha \left(u_\alpha^2 + 3p/\rho\right) - \sum_i c_{i,\alpha}^3 f^{\rm eq}_i.
\end{equation}
More advanced versions of the BGK operator are usually employed on the basis of either physical, for instance variable Prandtl number, or numerical arguments. Among those, the multiple relaxation time (MRT) formulation allows to relax the distribution function in a space of linearly independent base functions, typically moments of the distribution function. Owing to its improved stability and accuracy it has become quite popular in the lattice Boltzmann community. Here, we make use of a MRT in a space of modified Hermite central moments allowing for independent control over the bulk viscosity, as first described in~\cite{hosseini2022lattice}. At the difference of the classical Hermite polynomial space~\citep{hosseini2021central,huang2022simulation}, the trace and trace-free contributions to the second-order moments relax independently in this formulation:
\begin{equation}
    \Omega_i = \mathcal{T}^{-1}W\mathcal{T}(f^{\rm eq}_i - f_i) +  \Xi_i\label{eq:correc1}
\end{equation}
where $\mathcal{T}$ and $\mathcal{T}^{-1}$ are the moments transform tensor and its inverse, and $W$ is the diagonal tensor of relaxation rates. The moments transform tensor is defined via a set of modified central Hermite polynomials $\widetilde{\mathcal{H}}_n(\bm{c}_i)$ , which for the D3Q27 stencil are:

{%
    \begin{minipage}{\linewidth}
    \begin{multline}
    \widetilde{\mathcal{H}}_n \in \{ \widetilde{\mathcal{H}}_0, \widetilde{\mathcal{H}}_x, \widetilde{\mathcal{H}}_y, \widetilde{\mathcal{H}}_z, \widetilde{\mathcal{H}}_{xy}, \widetilde{\mathcal{H}}_{xz}, \widetilde{\mathcal{H}}_{yz}, \widetilde{\mathcal{H}}_{xx}-\widetilde{\mathcal{H}}_{yy}, \widetilde{\mathcal{H}}_{xx}-\widetilde{\mathcal{H}}_{zz}, \widetilde{\mathcal{H}}_{xx}+\widetilde{\mathcal{H}}_{yy}+\widetilde{\mathcal{H}}_{zz},\\ \widetilde{\mathcal{H}}_{xyz}, \widetilde{\mathcal{H}}_{xxy}, \widetilde{\mathcal{H}}_{xxz}, \widetilde{\mathcal{H}}_{xyy}, \widetilde{\mathcal{H}}_{yyz}, \widetilde{\mathcal{H}}_{xzz}, \widetilde{\mathcal{H}}_{yzz}, \widetilde{\mathcal{H}}_{xxyy}, \widetilde{\mathcal{H}}_{xxzz}, \widetilde{\mathcal{H}}_{yyzz}, \widetilde{\mathcal{H}}_{xxyz},\\ \widetilde{\mathcal{H}}_{xyyz},
    \widetilde{\mathcal{H}}_{xyzz}, \widetilde{\mathcal{H}}_{xyyzz}, \widetilde{\mathcal{H}}_{xxyzz}, \widetilde{\mathcal{H}}_{xyzz}, \widetilde{\mathcal{H}}_{xxyyzz} \}.
\end{multline}
    \end{minipage}%
}

For detailed expression of central Hermite polynomials we refer interested readers to \cite{Hosseini}.
Equilibrium moments based on these polynomials are the central Hermite coefficients $\widetilde{a}^{\rm eq}_n$, which for the equilibrium distribution function using D3Q27 are:\\
{%
    \begin{minipage}{\linewidth}
    \begin{multline}
    \widetilde{a}^{\rm eq}_n \in \{ \rho, 0, 0, 0, 0, 0, 0, 0, 0, 3(p - \rho c_s^2), 0, 0, 0, 0, 0, 0, 0, {(p - \rho c_s^2)}^2, \\ {(p - \rho c_s^2)}^2, {(p - \rho c_s^2)}^2, 0, 0, 0, 0, 0, 0, {(p - \rho c_s^2)}^3 \}.
\end{multline}
    \end{minipage}%
}
Setting $p=\rho c_s^2$, all equilibrium coefficients reduce to zero except at order zero. The relaxation frequency tensor $W$ for the D3Q27 stencil is defined as:\\
\colorbox{highlightcolor}{%
    \begin{minipage}{\linewidth}
    \begin{multline}
    W={\rm diag}(1, 1, 1, 1, \omega_s, \omega_s, \omega_s, \omega_s, \omega_s, \omega_b, \omega_g, \omega_g, \omega_g, \omega_g, \omega_g, \omega_g, \omega_g, \\ \omega_g, \omega_g, \omega_g, \omega_g, \omega_g, \omega_g, \omega_g, \omega_g, \omega_g, \omega_g),
\end{multline}
    \end{minipage}%
}
where the operator ${\rm diag}$ is defined as:
\begin{equation}
    {\rm diag}(\bm{A}) = (\bm{A}\otimes\bm{1})\circ \bm{I},
\end{equation}
for a given vector $\bm{A}$, with $\bm{1}$ a vector with elements 1, $\bm{I}$ the unitary tensor, and $\circ$ the Hadamard product, while $\omega_s$, $\omega_b$ and $\omega_g$ are the shear, bulk, and ghost-modes relaxation frequency, respectively. The quantity $\omega_b$ is related to the bulk viscosity, $\eta$, as~\cite{hosseini2022towards}:
\begin{equation}
    \omega_b = \frac{\delta t}{(\frac{2+D}{D}-\frac{\partial\ln p}{\partial\ln \rho})\frac{\eta}{p} + \delta t/2},
\end{equation}
while the shear modes relax with:
\begin{equation}
    \omega_s = \frac{\delta t}{\frac{\mu}{p} + \delta t/2},
\end{equation}
and $\omega_g=1$.
The correction term for the diagonal components of the third-order moments appearing in Eq.~\ref{eq:correc1} changes in the context of the multiple relaxation time collision operator into:
\begin{equation}\label{eq:correction_MRT}
    \Xi_i = \left(1-\frac{\omega_b\omega_s}{\omega_b+\omega_s}\right)\frac{w_i}{2c_s^4}\left(\bm{\nabla}\mathcal{H}_2\right):\Delta^{\rm eq}_3.
\end{equation}
The lattice Boltzmann model as introduced here allows for an independent bulk viscosity and Galilean-invariant dynamic and bulk viscosities. It is worth noting that all the LBM equations mentioned in this article are expressed in S.I. units and no non-dimensionalization has been used.
\subsection{Boundary conditions}
While a variety of different boundary conditions have been developed for LBM, in the context of the present study all solid boundaries can be simply modeled using the half-way bounce-back scheme, where missing distribution functions after the collision-streaming steps are computed as \citep{kruger2017lattice}:
\begin{equation}
    f_i\left(\bm{x},t+\delta t\right) = f^{*}_{\bar{i}}\left(\bm{x},t\right),
\end{equation}
where $f^{*}$ is the post-collision population (prior to streaming) and $\bar{i}$ is the index of the particle velocity opposite that of $i$. To get rid of any staircase approximation, the half-way bounce-back is supplemented with the advanced treatment for curved boundaries proposed in~\cite{bouzidi2001momentum}.\\
At a given boundary node $\bm{x}_f$, the missing incoming populations are computed as:
\begin{subequations}
	\begin{align}
		f_i(\bm{x}_f, t+\delta t) &= 2qf_{\bar{i}}(\bm{x}_f+\bm{c}_{\bar{i}}, t+\delta t)\nonumber \\  &+\left(1-2q\right)f_{\bar{i}}(\bm{x}_f, t+\delta t), \forall q<\frac{1}{2},\\
		f_i(\bm{x}_f, t+\delta t) &= \frac{1}{2q}f_{\bar{i}}(\bm{x}_f+\bm{c}_{\bar{i}}, t+\delta t)\nonumber \\  &+\frac{2q-1}{2q}f_{\bar{i}}(\bm{x}_f, t+\delta t), \forall q\geq\frac{1}{2},
		\end{align}
	\label{Eq:CE_moments_eq}
\end{subequations}
where $\bar{i}$ designates the direction opposite $i$ and $q$:
\begin{equation}
    q = \frac{\lvert\lvert \bm{x}_f - \bm{x}_s\lvert\lvert}{\lvert\lvert\bm{c}_i \lvert\lvert},
\end{equation}
with $\bm{x}_s$ denoting the wall position in direction $i$.\\
Other boundary conditions applied at the inlets and outlets, {i.e.}, constant velocity and/or constant pressure, are realized using the non-equilibrium extrapolation approach proposed in~\cite{zhao2002non}. In this method the missing populations at the boundary nodes are reconstructed as:
\begin{equation}
    f_{i}\left(\bm{x}_w,t+\delta t\right) = f^{\rm eq}_{i}\left(\bm{x}_w,t+\delta t\right) + f^{\rm neq}_{i}\left(\bm{x}_w,t+\delta t\right),
\end{equation}
where $\bm{x}_w$ is the position of the boundary node and $f^{\rm neq}$ is the non-equilibrium part of the distribution function. To be able to use this equation one needs both, density and velocity at the boundary node. One of these two variables is known via the boundary condition, and the other one is computed by interpolation from neighboring nodes. The non-equilibrium part of the distribution function is also computed via interpolation in space.\\
In addition to velocity and pressure boundary conditions, the zero-gradient outlet boundary condition was implemented following~\cite{kruger2017lattice}:
\begin{equation}
    f_i\left(\bm{x},t+\delta t\right) = f_i\left(\bm{x}-\bm{n}\delta x,t+\delta t\right),
\end{equation}
where $\bm{n}$ is the outward-pointing unit vector normal to the boundary surface.

One attractive feature of LBM is that, using such boundary conditions, there is no need for any complex grid adaption to resolve the structure of the packed bed. A simple, regular, equidistant grid is used in all simulations.
The employed resolution in time and space is discussed later on.
\section{Configuration considered in this study}
As explained previously, the same geometry, same size, same operating conditions have been considered for numerical simulations and experimental measurements, enabling direct comparisons discussed later. However, real experiments and diagnostics involve of course many more aspects needed for a practical realization. This is why the next sections describe separately particular features of the experimental bench and of the LBM simulations.

Practical applications of interest for our group involve typical particle Reynolds numbers $\mathrm{Re}_P$ between 100 and 1000, in most cases between 300 and 500. This is why these two particular values have been considered in the present study. Here the Reynolds number is defined as:
\begin{equation}
    \mathrm{\mathrm{Re}_P} = \frac{\rho d_p v_{int}}{\mu},
\end{equation}
where $d_p$ is particle diameter, $\mu$ is the dynamic viscosity of air for ambient pressure and temperature ($p = 1~\mathrm{bar}$ and $T$ = 20 °C), $\mu = 1.822 \cdot 10^{-5} \, \mathrm{Pa \cdot s}$ and $\rho$ is the air density under the same conditions, $\rho = 1.204 \, \mathrm{kg/m^3}$. The interstitial velocity $v_{int}$ is defined as:
\begin{equation}
    v_{\mathrm{int}} = \frac{v_s}{\varphi} = \frac{Q}{A\cdot \varphi},
\end{equation}
where $v_s$ is the superficial velocity based on flow-rate and cross-section of the reactor $A = 0.132\times 0.132~\mathrm{m^2}$ and $\varphi$ the porosity of the packed bed, here $\varphi = 0.32$. $Q$ is the volume flow rate, here taken to be  $Q = 38.2$ and $63.6~\mathrm{l/min}$ for Reynolds 300 and 500, respectively.
\label{sec:set_up}
\subsection{Experimental setup}

Advanced imaging technologies, such as Radar Tomography or Magnetic Resonance Imaging are considered in companion projects of this study. They appear very promising, but are extremely complex and do not provide a very high resolution in space and/or time~\citep{Tropea,TropeaMRI,Poelma.2020}. The experimental experience of our own group is focused on optical flow measurements. Regarding velocity and fluctuations, Particle Image Velocimetry (PIV) is currently the gold standard. However, PIV is only possible with sufficient optical access. For almost all published studies using optical measurement techniques in packed beds, Refractive Index Matching (RIM) has been used, employing a liquid with the same refraction index as the (transparent) bed particles to enable PIV~\citep{hassan2008flow,Larsson.2018,wood.2015} or adjusting the solid material to the refractive index of the working fluid~\citep{harshani.2016}. 
\cite{hassan2008flow} applied Particle Tracking Velocimetry (PTV) and high-speed PIV to investigate the flow in a vertical packed bed of randomly arranged PMMA spheres with a diameter of 4.7 mm for particle Reynolds numbers of 80 to 500 and characterised the vortex formation at the pore level scale. Two dimensional flow fields of laminar flows with a particle Reynolds number in the range of 5 to 600 through a porous bed of glass spheres with a diameter of 15~mm where obtained using PIV by~\cite{wood.2015}. The imaging system was used to identify the bead centre location allowing to generate the exact geometry of the randomly assembled bed, used for direct comparison to DNS simulations. \cite{Larsson.2018} applied tomographic PIV to the horizontal flow through a thin porous bed consisting of vertical cylinders, which allowed a visualization of three-dimensional vortex structures and a determination of velocity magnitude for particle Reynolds numbers from 45 to 950.\\
However, the applications considered in the present study all involve gas flows, and it is almost impossible to keep identical all non-dimensional parameters characterizing momentum, heat, and mass transfer when switching to a liquid using flow similarity. As a consequence, Refractive Index Matching cannot be used, and alternatives must be found. In 2D porous media, innovative methods have been introduced for visualizing interstitial pore fluid flow using fluorescence imaging and high-speed particle tracking~\citep{li2022visualization}. However, they cannot be directly applied to real, 3D configurations. To the best of our knowledge, no systematic PIV measurements have been carried out for gaseous flows in packed beds yet. A corresponding procedure is currently being developed in our group to avoid Refractive Index Matching. After first using a simple calibration to correct resulting optical distortions, \cite{Martins.2018} applied a ray-tracing based correction to measure by PIV a jet flow behind one to three transparent spheres. A current project of our group considers further improvements of this procedure, enabling PIV within beds constituted of optically-transparent spheres with gaseous flow. This leads to many challenges discussed in separate publications~\citep{velten2022a,Ebert.2022}.\\

\subsubsection{Description and instrumentation}
\begin{figure}[h]
	\centering
	\subfigure[]{\includegraphics[width=4cm,keepaspectratio]{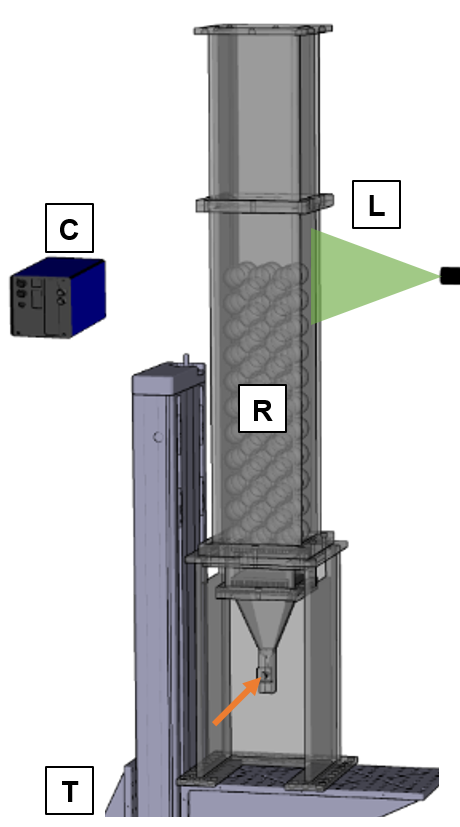}}
	\subfigure[]{\includegraphics[width=3.5cm,keepaspectratio]{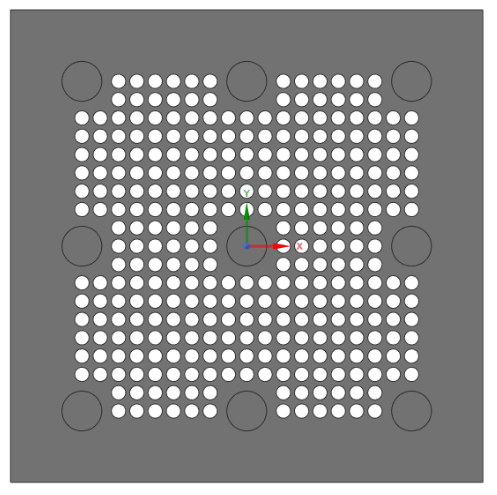}}
    \subfigure[]{\includegraphics[width=4cm,keepaspectratio]{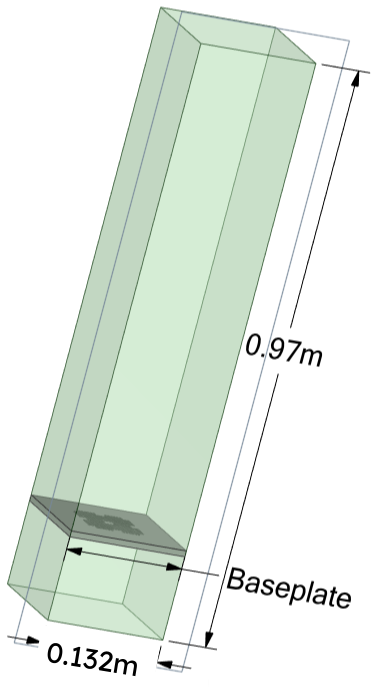}}
	\caption{(a) Scheme of the experimental setup with the camera (C) imaging the flow within or above the bulk reactor (R) containing the BCC packed bed, the whole system being installed on a 3D traversing unit (T) and illuminated by a laser light sheet (L). The orange arrow shows the gas inlet. (b) Top view of the reactor baseplate, showing with grey circles the nine bearings holding the spheres of the first layer at their target positions. (c) Numerical model and dimensions of the reactor used for the simulations. The packing is not shown here for increased readability (see later Fig.~\ref{MeasurementZones} for a representation with BCC packing).}
	\label{set_up}
\end{figure}

The gas flow field through a packed bed with 21 layers of spheres (the choice of this particular number will be discussed later) assembled in a body-centered-cubic (BCC) packing is investigated by standard PIV. The main components of the experiment, consisting of the bulk reactor mounted on a 3D-traversing unit, the laser light sheet (in green), and the PIV camera are shown in Fig.~\ref{set_up}(left). Liquid Di-Ethyl-Sebacat (DEHS) tracer particles provided by a liquid nebulizer (Type AGF 10.0, Palas GmbH) enter together with the air adjusted by a mass-flow controller (Bronkhorst D-6371; max. 500 l/min air) through the gas inlet at the bottom (orange arrow in Fig.~\ref{set_up}, left). The light sheet is created by a Nd:YAG PIV laser (Quantel Q-smart Twins 850) with a maximum energy of 380 mJ/pulse operated at a wavelength of 532 nm. In what follows, PIV measurements will be discussed both on the freeboard, just above the packed bed, as well as in the interstices between the particles. For the measurement above the packing, an Imager LX 8M camera equipped with a Nikkor 35 mm f/2D Nikon lens has been used. Within the packed bed, an Imager Intense camera equipped with a Micro-Nikkor 105 mm f/2.8D Nikon lens has been employed. Image acquisition took place at a frequency of 2.5 Hz. Different recording times were applied as function of the particle Reynolds number, acquiring 300 double-frame images for $\mathrm{Re}_P = 300$ and 1000 for $\mathrm{Re}_P = 500$. All PIV evaluations were carried out with the commercial software Davis 8.4 from LaVision GmbH. A standard cross-correlation method (multi-pass, decreasing size from 64 x 64 to 32 x 32 pixels with 50\% overlap) has been employed to calculate the two velocity components within the laser sheet, leading to two-dimensional vector fields. The 3D traversing unit allowed for an accurate adjustment of the measurement position without moving any optical component.

\subsubsection{Details concerning the experimental packing geometry}
The flow through the reactor shown in Fig.~\ref{set_up}(left) involves three distinct parts: first, the inlet zone, until the baseplate; then, the BCC packing itself; finally, an extended outlet zone with no particles but side walls, to minimize any uncontrolled influence from the surroundings. The reactor has been designed to yield as much as possible symmetric, stable flow conditions, since it involves a fully symmetric geometry and constant inlet flow-rate. Additionally, a diffuser with a combination of honeycombs and $4~\mathrm{mm}$ glass spheres is used to homogenise the flow in the inlet zone and distribute it equally over the whole cross-section of the baseplate shown in Fig.~\ref{set_up}(b). A pattern of 312 regularly-arranged holes (with a diameter of $d_h= 4~\mathrm{mm}$ each) controls the flow entering the packing. No holes are placed near the vertical walls in order to reduce wall channeling. The nine spheres building the first particle layer are fixed in proper position by small cavities (maximal depth of $0.8$~mm), shown as gray circles in Fig.~\ref{set_up}(b).\\

The BCC packing itself consists of 139 high-precision polypropylene spheres with a diameter of 40 mm (tolerance $150~\mu$m, from Ningyang Xinxin Stainless Steel Ball Manufacture Co., Ltd.), involving a maximum of 11 so-called "full layers" (3 x 3 particles, touching the vertical walls of the container) and 10 "weak layers" in-between (only 2 x 2 particles, without contact to the walls). The resulting height of the BCC packing consisting of 21 layers is $0.51$ m starting from the baseplate. The distance between two opposite vertical walls is $0.132$~m (square cross-section). Including additionally the outlet zone, the vertical walls of the container extend up to $0.92$~m above the ground plate.

Note that, in order to avoid any micro-movement or vibration of the spheres when activating the traversing unit, all spheres have been glued together by a melting adhesive, creating a local bridge at the contact points between the particles. This leads to minute deviations from the ideal BCC packing. Considering that these slight modifications cannot be characterized with sufficient accuracy -- and have, therefore, not been taken into account in the simulations -- all comparisons shown later on will assume that the experimentally-employed BCC packing corresponds to an unperturbed, ideal configuration.

\subsection{Simulation setup}
\subsubsection{Numerical domain}
In the LBM simulations, a slightly simplified version of the geometry is used, as shown in Fig.~\ref{set_up}(c). Note that the BCC packing is not shown in this figure for a better readability. All important features controlling the flow are kept identical to the experimental conditions. The geometry consists of a 3D domain with a total height of $0.97$~m, and a constant cross-section of $0.132\times 0.132$~m$^2$ (identical to the inner dimensions of the experimentally-employed container). The same baseplate shown in Fig.~\ref{set_up}(b) is placed 0.14~m downstream of the inlet plane in the simulations. The boundary condition at this inlet plane corresponds to a fully-homogeneous flow, mimicking the action of diffuser, honeycombs, and glass beads used to homogenize the flow in the experiments. After the baseplate (10 mm thickness), an ideal BCC packing is implemented (not shown in Fig.~\ref{set_up}c), involving up to 21 layers -- as in the experiments. After reaching the top-end of the packed bed, the numerical domain is extended by $0.31\,\rm{m}$ in streamwise direction until the outlet boundary condition (BC). Preliminary tests have shown that this is sufficient to avoid any undue influence of the outlet BC on the comparisons discussed later on (though it is somewhat shorter than in the experiments). When implementing the highest number of layers (21), the total height of the numerical domain is $0.97\,\rm{m}$.

The inlet boundary condition is set to a constant volume flow-rate boundary condition implemented via a modified form of the half-way bounce-back method~\citep{kruger2017lattice}. At the outlet, a constant-pressure boundary condition enforced through the non-equilibrium extrapolation method~\citep{zhao2002non} is used, and no-slip boundary conditions are implemented along the vertical walls. The bounce-back method with curved treatment extension, detailed in previous sections, is used to enforce the no-slip condition.

The dimensions of the 3D computational domain are $0.132\times 0.132\times0.97\,\rm{m^3}$. After preliminary tests, the grid resolution used for all simulations corresponds to $\delta x=5 \times10^{-4} \rm{m}$ (same in all directions) so that each hole in the baseplate is resolved by 8 grid points. This leads finally to a grid with $\approx 135$~Mio. grid points. The time-step sizes are set to $\delta t=6.88\times10^{-5}\, \rm{s}$ for $\mathrm{Re}_P=300$, and $\delta t=4.13\times10^{-5}\, \rm{s}$ for $\mathrm{Re}_P = 500$. This choice of time-step size guarantees that the maximum non-dimensional velocity in the domain (corresponding to the Courant-Friedrichs-Lewy condition, CFL) below $0.05$ guaranteeing minimal compressibility effects in the obtained solution. Note that in the context of the present study, we consistently use the convective CFL number based on maximal fluid velocity, defined as:
\begin{equation}
    {\rm CFL} = \frac{\|\bm{u}\|_{\rm max}\delta t}{\delta x}
\end{equation}
All simulations have been carried out for a total of $12\ t_c$ in order to reach steady-state conditions (possibly with remaining, intrinsic fluctuations, as discussed later on), where $t_c$ is the flow-through time defined as:
\begin{equation}
 t_c = \frac{L_c}{v_s},
\end{equation}
where the characteristic speed is the superficial velocity and characteristic length $L_c$ is the length of the entire flow domain in streamwise direction. Time-averaging is only performed on data from the second half of the simulation (6\textendash 12 $t_c$) to compute the average flow fields (and associated fluctuations).

\subsection{Locations used for comparisons and analysis}
\begin{figure}[h!]
	\centering\includegraphics[width=\textwidth,keepaspectratio]{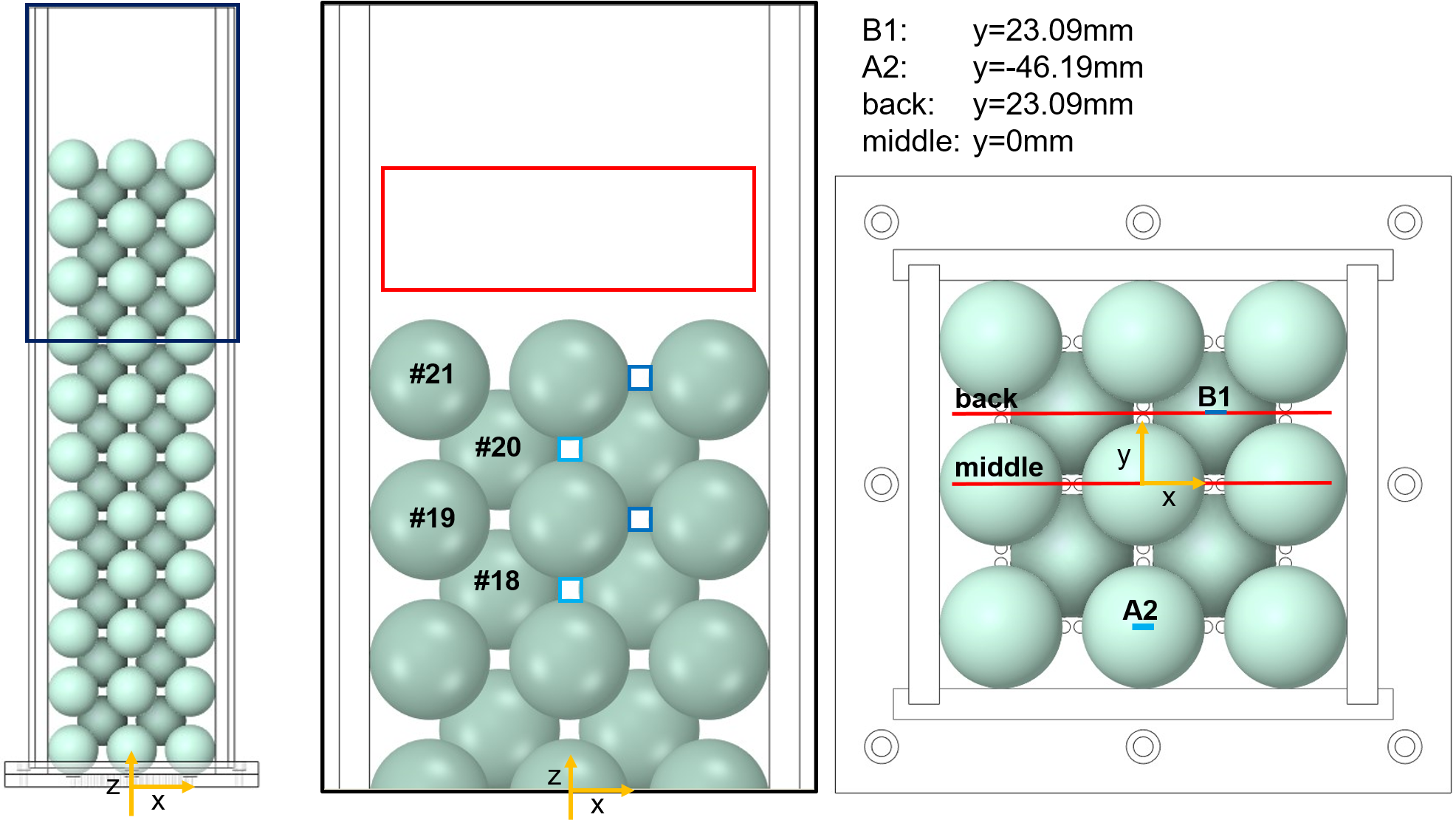}
	\caption{Front view of the packed bed reactor with 21 layers of spheres in a BCC packing (left). A zoom of the region of interest for the later comparisons (layers \#18 to \#21, and freeboard above the bed) is marked with a blue rectangle in the left figure, and shown in the central image. In this central view, the regions where the flow fields were measured by PIV and will be compared to simulations are marked: the red rectangle corresponds to the freeboard measurements above the bed (Config. 2), while the small regions delimited by dark blue lines show the interstitial measurements (Config. 1) within the full layers (\#19 and \#21), and those in light blue corresponding to the weak layers (\#18 and \#20). Finally, to clarify the corresponding positions in cross-direction (along the $y$-axis), a top view is added on the right: red lines show measurement planes above the bed, while short blue lines correspond to measurements within the packed bed (B1 and A2).}
	\label{MeasurementZones}
\end{figure}

PIV measurements for $\mathrm{Re}_P = 300$ or $500$ have been carried out for two different configurations using the same experimental setup: Configuration 1, within selected inter-particle spaces for layer \#18, 19, 20, or 21, delivering useful information within the packed bed for locations that are optically accessible from outside, and Configuration 2, just above the BCC packing involving 21 layers of spheres (sometimes also called freeboard measurements in what follows).
All these measurement zones are illustrated and described in Fig.~\ref{MeasurementZones}. Since the corresponding information is not easy to convey, but important for the later analysis, the reader is invited to spend sufficient time on this figure, and come back to it later if necessary. 

While all experimental results discussed in this article have been obtained for a total number of 21 layers, the simulations have been repeated as well in a similar procedure for 15, 17, 19, and 21 total layers in the bed.
When discussing corresponding positions in an absolute manner, the same Cartesian coordinate system is always used, as shown in Fig.~\ref{MeasurementZones}. The $x$ and $y$ axes define horizontal cross-sections, with the origin at the geometrical center of the reactor, $x$ called horizontal direction and $y$ called depth. The $z$-axis corresponds to the vertical, streamwise direction (also called height).

\section{Comparisons with experimental data and cross-validation}
The numerical results obtained with lattice Boltzmann are now compared to the PIV data based on the two configurations described previously (within the bed, above the bed). This procedure allows for a cross-validation, a good agreement demonstrating the accuracy of the simulation procedure, while using at the same time the additional details provided by LBM to better understand experimental observations. The employed scales and domain dimensions are always the same for experimental and numerical data, enabling a direct comparison.

\subsection{Configuration 1: within the bed}
\label{Inside}
First comparisons concern the flow within the packed bed. Only small optically accessible regions of around 7 x 7~mm can be used for PIV between the spheres, as shown previously in Fig.~\ref{MeasurementZones}. They are now considered for direct comparisons with simulation results. 
Two different positions (B1 and A2, as plotted in Fig.~\ref{MeasurementZones}) are involved: B1, situated in the same $y$ position as the back-plane discussed in the next section, is farther from the back wall of the container, whereas A2 is close (only half a sphere diameter) to the front wall.

 \begin{figure}[h!]
	\centering
	\includegraphics[width=\textwidth,keepaspectratio]{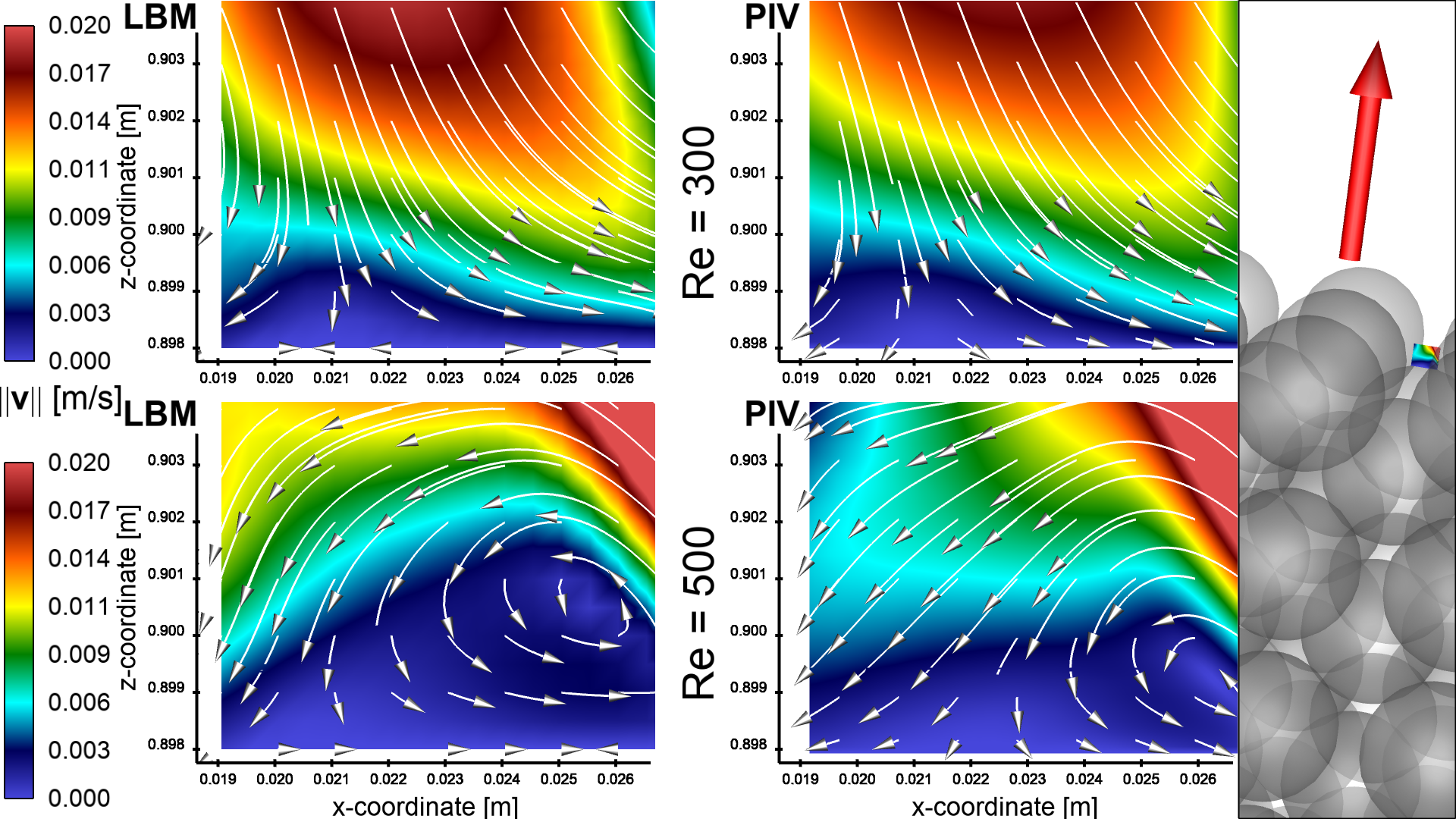}
	\caption{Comparison of the averaged in-plane velocity magnitude from LBM (left) and PIV (right) with an overlay showing streamlines for position B1 in layer \#21 for $\mathrm{Re}_P = 300$ (top row) and $\mathrm{Re}_P = 500$ (bottom row). For a better visibility, only every 9th vector (PIV) and every 2nd grid value (LBM) are displayed.}
	\label{B1_L21}
\end{figure}
 \begin{figure}
	\centering
	\includegraphics[width=\textwidth,keepaspectratio]{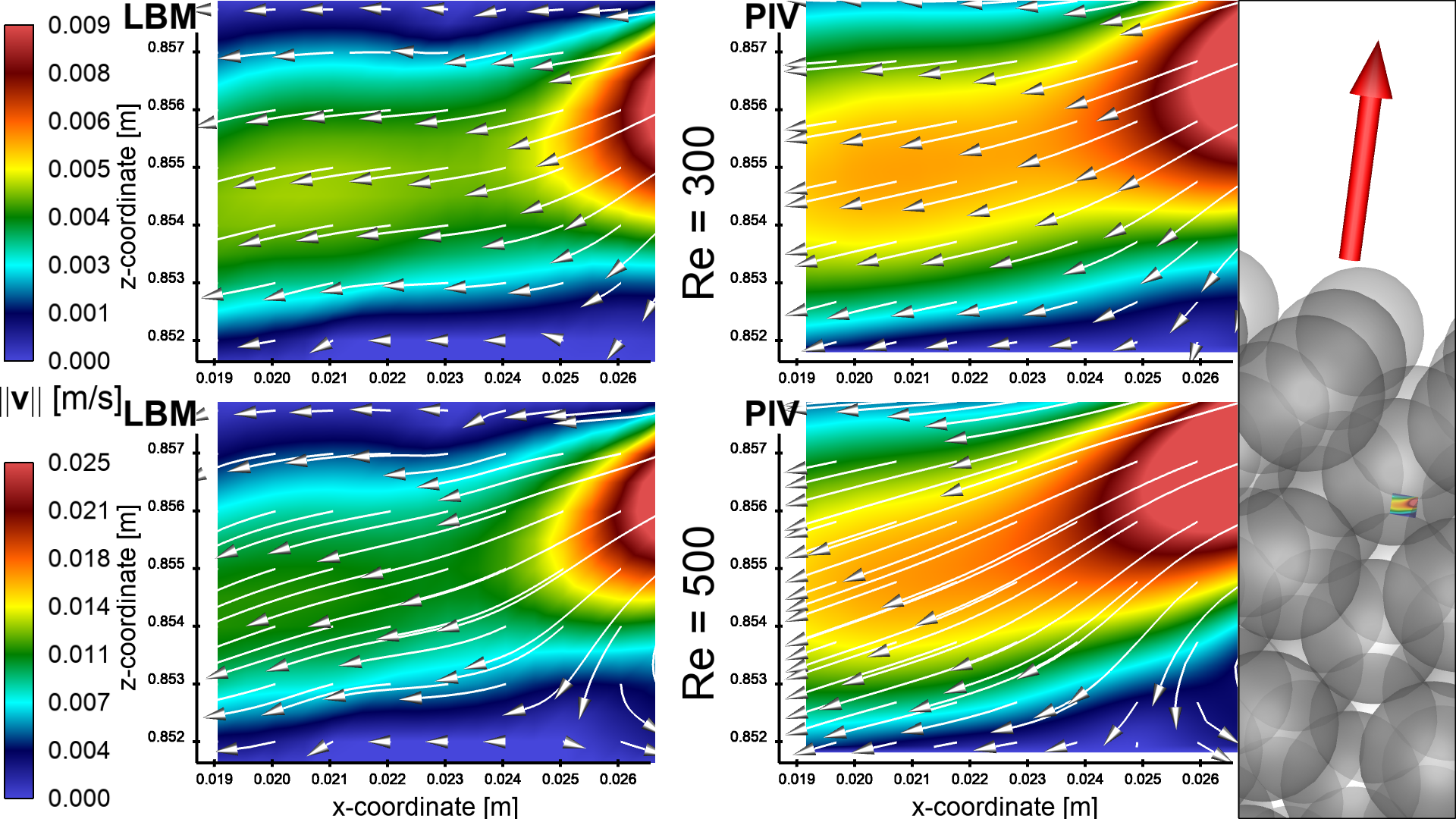}
	\caption{Comparison of the averaged in-plane velocity magnitude from LBM (left) and PIV (right) with an overlay showing streamlines for position B1 in layer \#19 for $\mathrm{Re}_P = 300$ (top row) and $\mathrm{Re}_P = 500$ (bottom row). Please note that different color bars are used for the two different Reynolds numbers -- but are of course the same for numerical and experimental results. For a better visibility, only every 9th vector (PIV) and every 2nd grid value (LBM) are displayed.}
	\label{B1_L19}
\end{figure}
Figure~\ref{B1_L21} shows the flow fields in the interstice B1 for layer \#21. This is the last inter-particle space before the flow gets released to the surface of the bed. In Fig.~\ref{B1_L19} the same interstice, but now in layer \#19 (deeper within the bed) is presented. The same dimensions and color scales are used for LBM and PIV, so that these visualisations allow a direct comparison between numerical and experimental results for both Reynolds numbers, $\mathrm{Re}_P = 300$ and $\mathrm{Re}_P=500$. All plots show the in-plane average velocity magnitudes involving the two velocity components in $x$ and $z$ directions (since the $y$-component could not be measured by PIV), with an overlay of streamlines showing also flow directions.

\begin{figure}
	\centering
	\includegraphics[width=\textwidth,keepaspectratio]{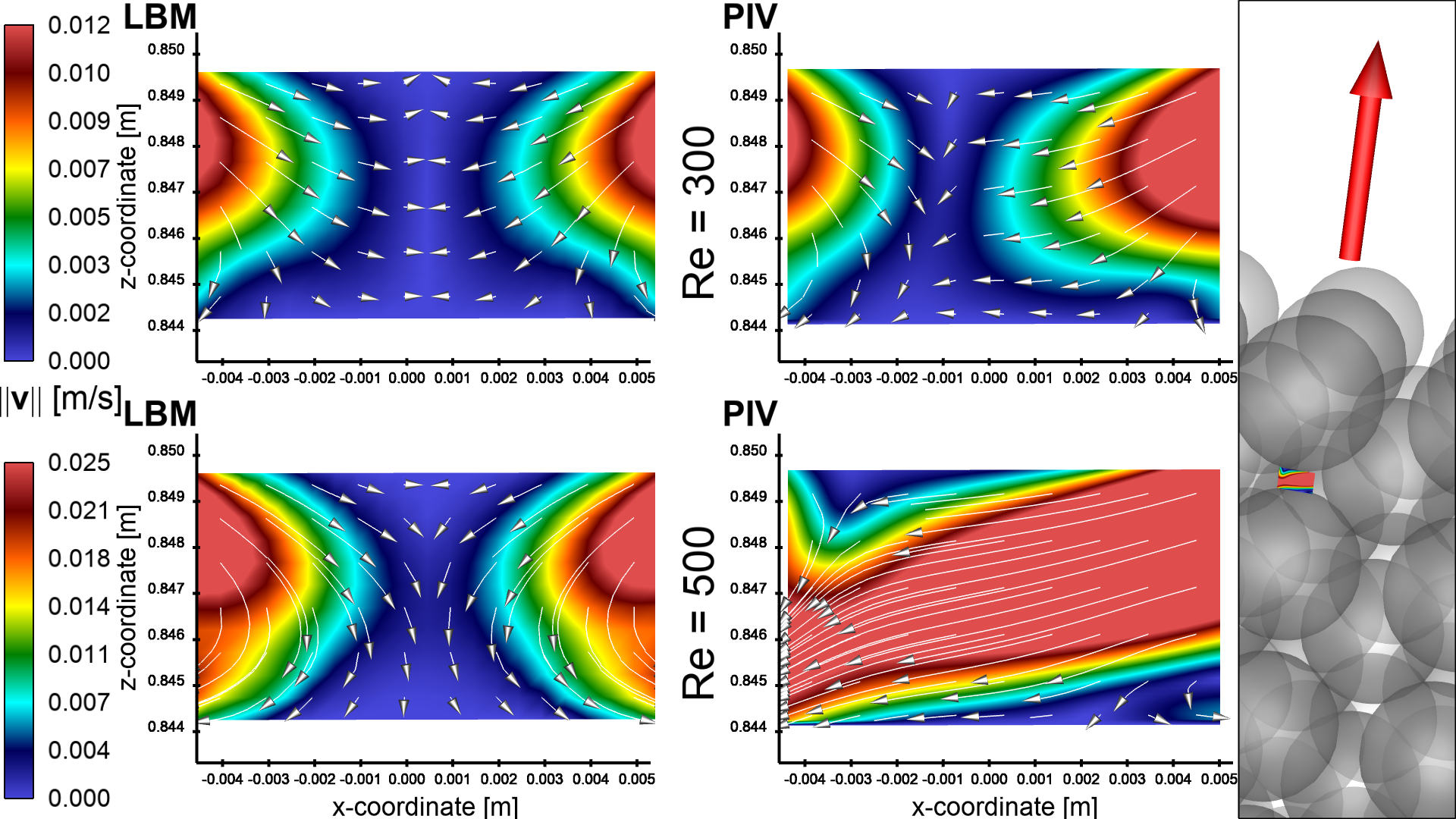}
	\caption{Comparison of the averaged in-plane velocity magnitude from LBM (left) and PIV (right) with an overlay showing streamlines for position A2 in layer \#20 for $\mathrm{Re}_P = 300$ (top row) and $\mathrm{Re}_P = 500$ (bottom row). Please note that different color bars are used for the two different Reynolds numbers -- but are of course the same for numerical and experimental results. For a better visibility, only every 9th vector (PIV) and every 2nd grid value (LBM) are displayed.}
	\label{A2_L20}
 \end{figure}

\begin{figure}
	\centering
	\includegraphics[width=\textwidth,keepaspectratio]{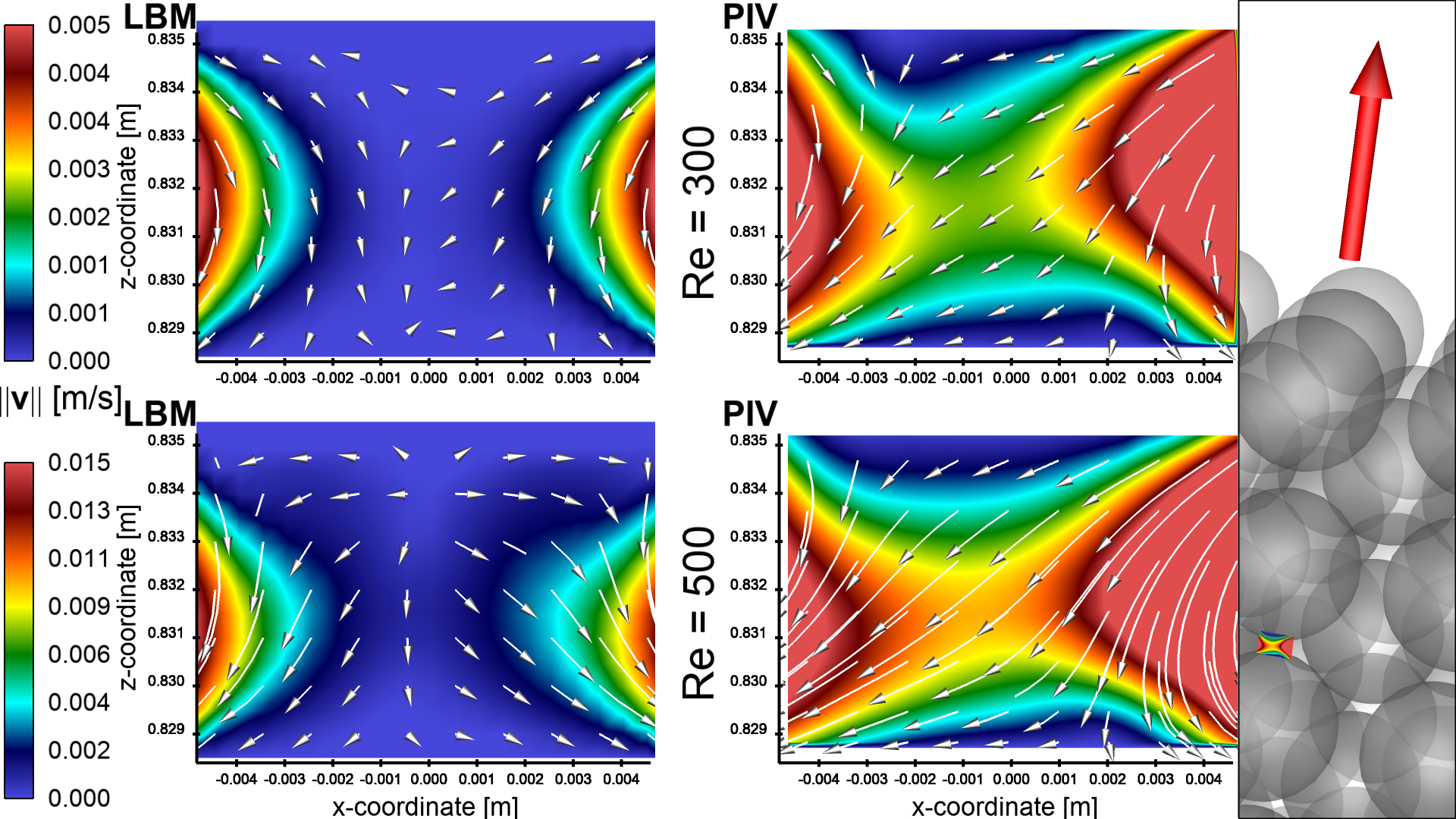}
	\caption{Comparison of the averaged in-plane velocity magnitude from LBM (left) and PIV (right) with an overlay showing streamlines for position A2 in layer \#18 for $\mathrm{Re}_P = 300$ (top row) and $\mathrm{Re}_P = 500$ (bottom row). Please note that different color bars are used for the two different Reynolds numbers -- but are of course the same for numerical and experimental results. For a better visibility, only every 9th vector (PIV) and every 2nd grid value (LBM) are displayed.}
	\label{A2_L18}
\end{figure}

Because of the flow getting released to the free surface, the measured fields in the top layer (\#21, Fig.~\ref{B1_L21}) are very different from those deeper inside the bed (\#19, Fig.~\ref{B1_L19}); even the structures are completely different. While layer \#21 is dominated by a back-flow at $\mathrm{Re}_P = 300$, or a large recirculation zone at $\mathrm{Re}_P = 500$, the results within layer \#19 reveal a strongly horizontal flow from right to left. Since in-plane velocity magnitudes increase only slightly with particle Reynolds numbers near the free surface, the same color scale has been used. On the other hand, the peak velocity becomes much higher when increasing Re$_P$ within the bed (layer \#19) so that different color scales had to be employed in Fig.~\ref{B1_L19}.\\
For both positions and both values of $\mathrm{Re}_P$ a very good agreement between numerical predictions and experimental measurements is observed. Only very slight shifts in position are sometimes observed. The differences are a bit larger for $\mathrm{Re}_P = 500$, perhaps due to larger fluctuations (as discussed later in Sec. \ref{Fluctuations}).\\
The in-plane velocity magnitudes at position A2 (see again Fig.~\ref{MeasurementZones}) situated close to the reactor walls are now compared. Results from LBM and PIV for $\mathrm{Re}_P = 300$ and $\mathrm{Re}_P=500$ are shown in Figs.~\ref{A2_L20} and \ref{A2_L18} for layers \#20 (closer to the bed surface) and \#18 (deeper in the bed), respectively. The position of A2 in the horizontal symmetry plane of the bed ($x=0$) should yield symmetrical flow conditions, which is overall captured by the results, in particular for the numerical data. The flow field in the inter-particle space at position A2 is characterized by a strong vortex pair, which is found both in LBM and in PIV. The velocity magnitudes are always higher within layer \#20, which is the last weak layer before the flow gets released to the surface, in comparison to the deeper layer inside the bed (layer \#18); note the different color scales used in Figs.~\ref{A2_L20} and \ref{A2_L18}. The most pronounced difference between PIV and LBM is visible for $\mathrm{Re}_P = 500$ in Fig.~\ref{A2_L20} (bottom row). Here, the experimental data does not show fully symmetrical conditions and the flow structures are shifted to the upper left and the lower right corners, revealing perhaps a minute misalignment in the placement of the spheres for the experiments.\\
Overall, slightly larger differences are observed for position A2 (closer to the container wall) compared to position B1. Close to the A2 location the local porosity of the bed is higher because of the missing half-sphere at the wall in each weak layer. This, in connection with very slight modifications compared to an ideal BCC packing (as discussed in Section~\ref{sec:discus}) might explain the non-symmetrical flow features measured by PIV, as well as differences compared to numerical results. Obviously, LBM simulations assume a perfect BCC packing and only contact points between the spheres/with the walls.


\subsection{Configuration 2: just above the bed} \label{FlowAbove21Layers}
\begin{figure}[h!]
  \centering
   \vspace{0.5cm}
   \hspace{-0.5cm}
 {\includegraphics[width=10cm,keepaspectratio]{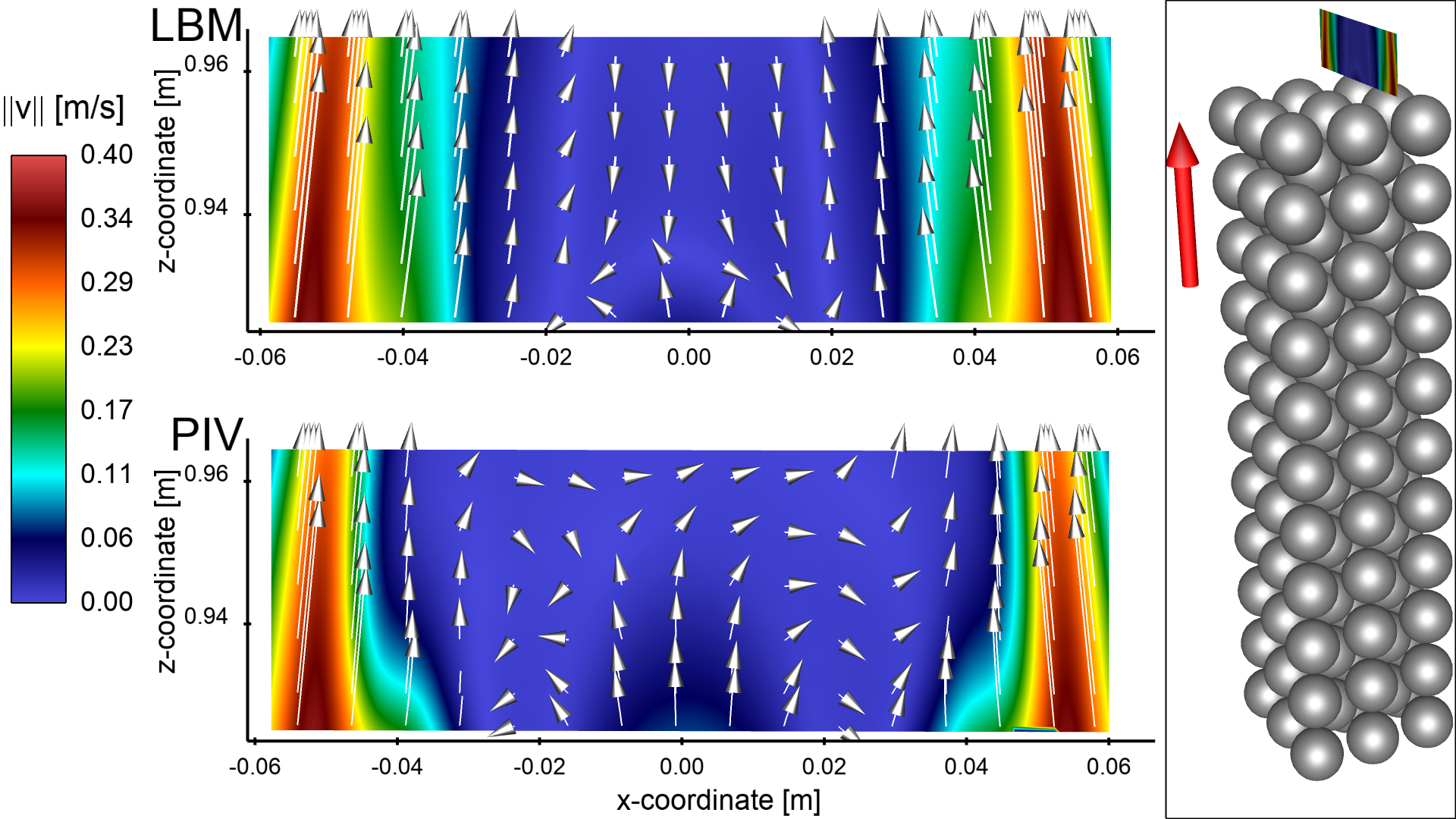}}
	\caption{Comparison of numerical data (LBM, top) and experimental results (PIV, bottom) for $\mathrm{Re}_P = 300$ at the back position above the bed concerning the averaged in-plane velocity magnitudes, with the corresponding vector field shown as an overlay to illustrate the flow direction. For a better visibility, only every 10th vector (PIV) and 15th grid value (LBM) vector are displayed -- the full resolution is much higher.\label{Re300_L21_back}}
\end{figure}

\begin{figure}[h!]
  \centering
   \hspace{-0.5cm}
{\includegraphics[width=10cm,keepaspectratio]
  {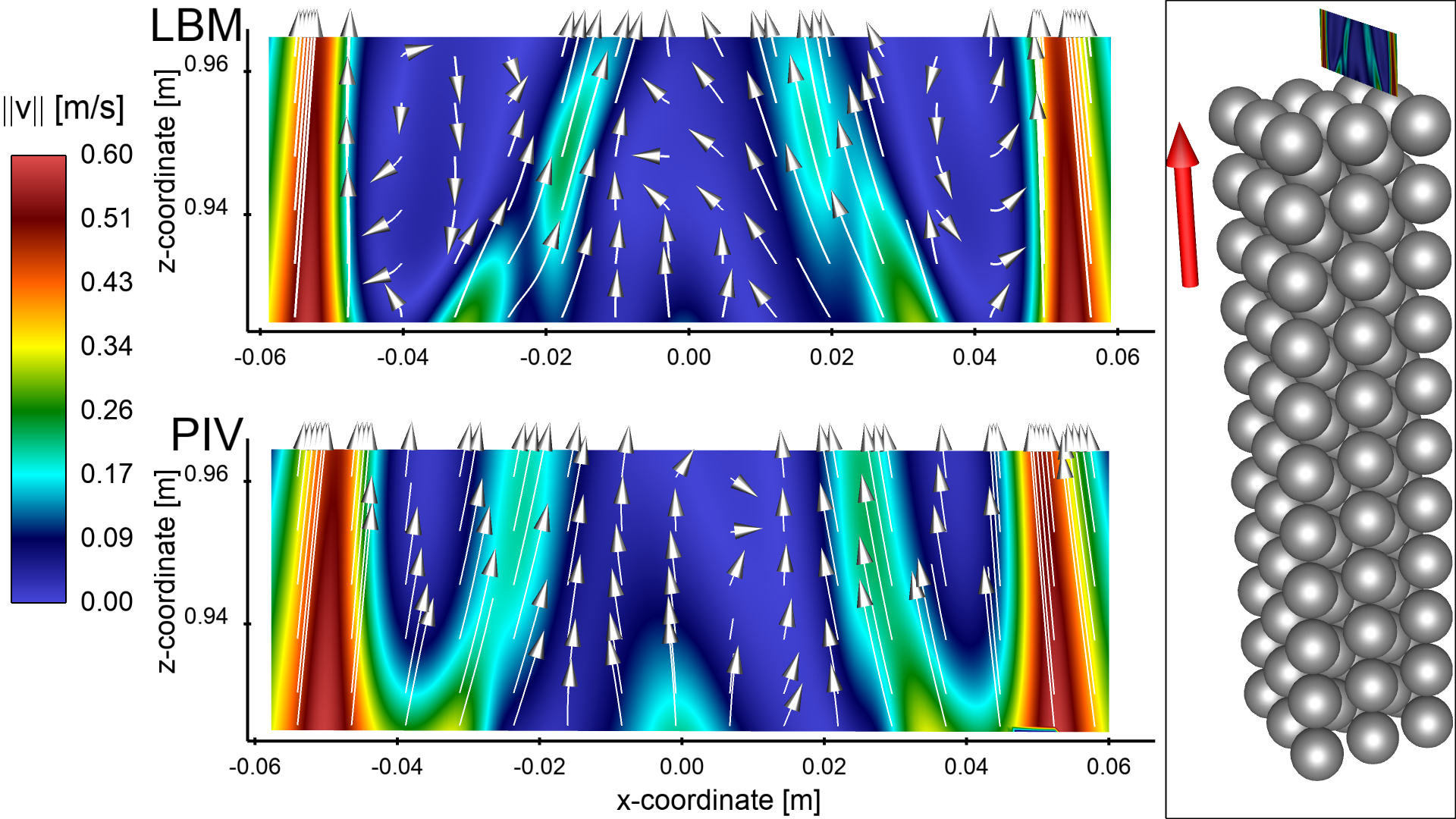}}
	\caption{Comparison of numerical data (LBM, top) and experimental results (PIV, bottom) for $\mathrm{Re}_P = 500$ at the back position above the bed concerning the averaged in-plane velocity magnitudes, with the corresponding vector field shown as an overlay to illustrate the flow direction. For a better visibility, only every 10th vector (PIV) and 15th grid value (LBM) vector are displayed.\label{Re500_L21_back}}
\end{figure}

\begin{figure}[h!]
  \centering
   \hspace{-0.5cm}
 {\includegraphics[width=10cm,keepaspectratio]{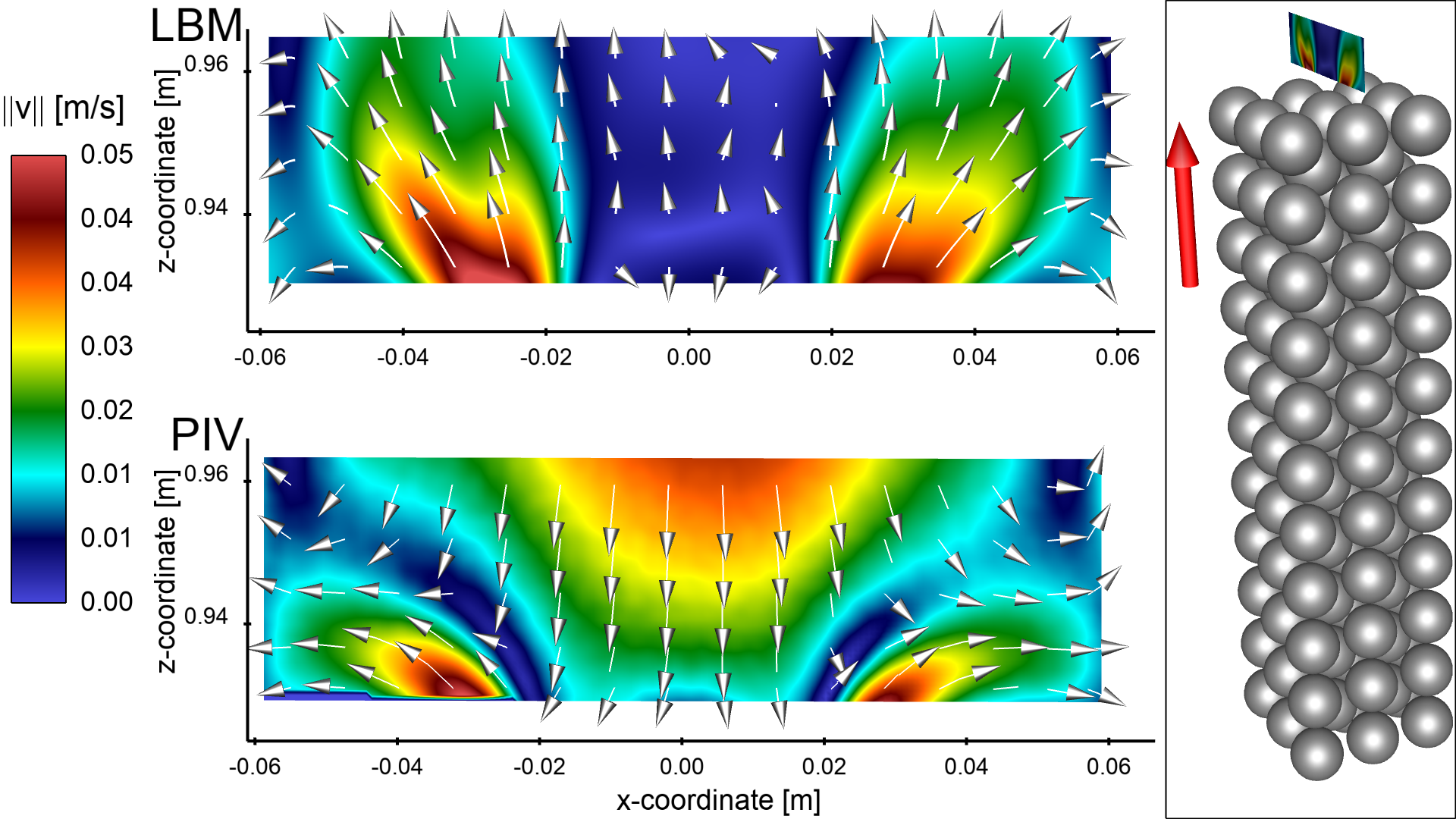}}
	\caption{Comparison of numerical data (LBM, top) and experimental results (PIV, bottom) for $\mathrm{Re}_P = 300$ at the middle position above the bed concerning the averaged in-plane velocity magnitudes, with the corresponding vector field shown as an overlay to illustrate the flow direction. For a better visibility, only every 10th vector (PIV) and 15th grid value (LBM) vector are displayed.\label{Re300_L21_middle}}
\end{figure}

\begin{figure}[h!]
  \centering
   \hspace{-0.5cm}
{\includegraphics[width=10cm,keepaspectratio]
{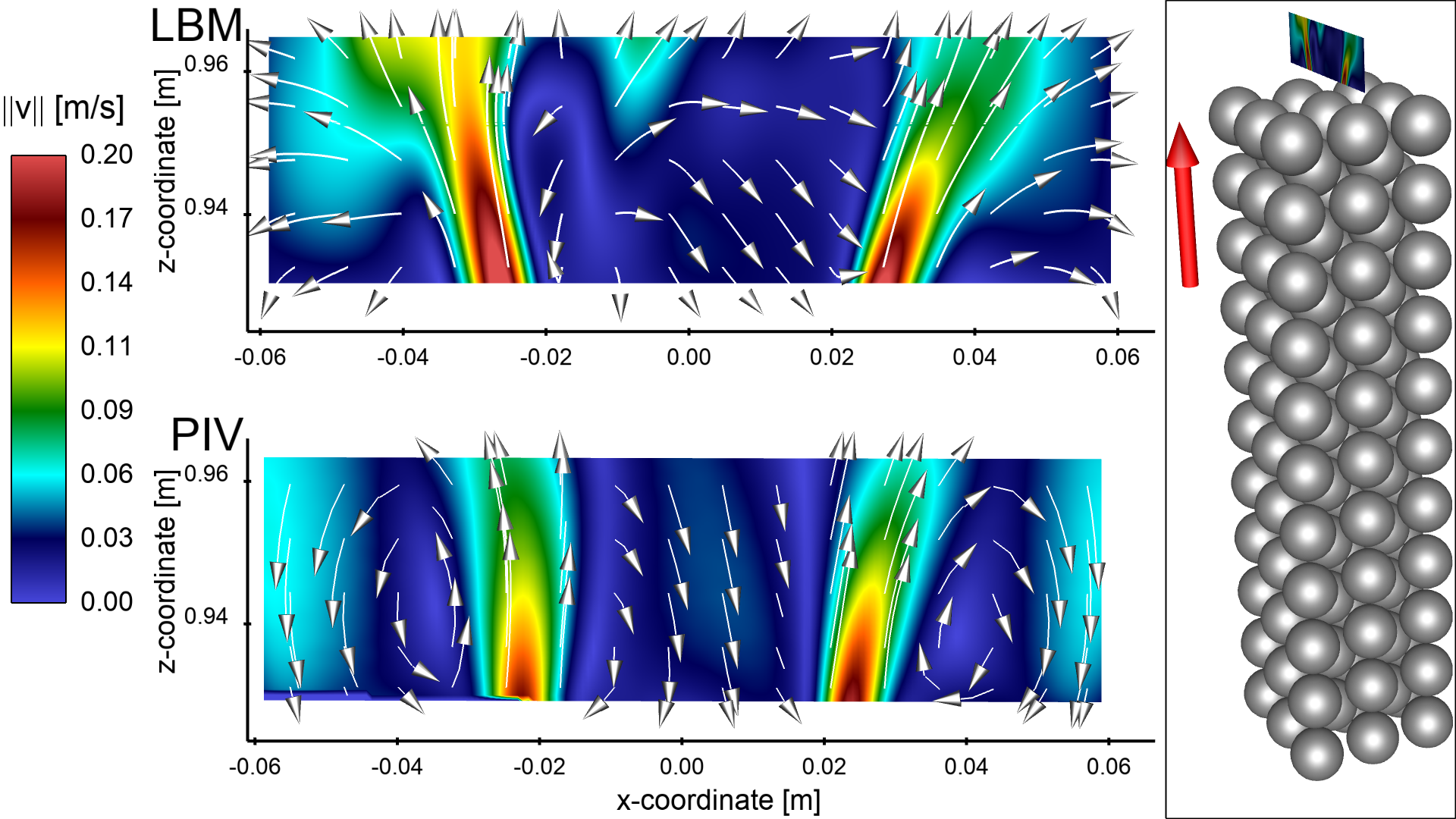}}
	\caption{Comparison of numerical data (LBM, top) and experimental results (PIV, bottom) for $\mathrm{Re}_P = 500$ at the middle position above the bed concerning the averaged in-plane velocity magnitudes, with the corresponding vector field shown as an overlay to illustrate the flow direction. For a better visibility, only every 10th vector (PIV) and 15th grid value (LBM) vector are displayed.\label{Re500_L21_middle}}
\end{figure}

The flow field downstream of the surface of the packed bed was measured within the red rectangle shown in Fig.~\ref{MeasurementZones} (center). The resulting flow fields obtained experimentally and numerically for $\mathrm{Re}_P = 300$ and $500$ are compared for two different measurement positions: back (Figs.~\ref{Re300_L21_back} and Fig.~\ref{Re500_L21_back}) and middle (Figs.~\ref{Re300_L21_middle} and Fig.~\ref{Re500_L21_middle}). The corresponding positions have been illustrated in Fig.~\ref{MeasurementZones}. All the results shown start $10~\mathrm{mm}$ above the bed. Assuming a stable flow, symmetric results are expected due to the symmetric geometry. Indeed, symmetric flow features are mostly observed in Figs.~\ref{Re300_L21_back} to \ref{Re500_L21_middle}.

The back position (Figs.~\ref{Re300_L21_back} and \ref{Re500_L21_back}) yields a flow field that is highly influenced by wall channeling effects, due to the missing half-sphere close to the wall in all weak layers. Therefore, two high-speed jets appear close to the walls. They are found for both Reynolds numbers, and are captured similarly by simulations and experiments. Comparing the numerical results with the experimental ones, a very good agreement is found for $\mathrm{Re}_P=300$. However, the LBM simulations overestimate slightly the height and the width of the jets, inducing somewhat smaller recirculation zones in the central part of the domain. The flow at $\mathrm{Re}_P = 500$ is again dominated by the high-speed jets near the side walls. Experiments and simulations reveal two additional, but less pronounced oblique jets at positions where, in this investigation plane, the flow is not blocked by the underlying layer of spheres. The agreement between PIV and LBM is again good at this location.

The next comparisons involve the middle position (Figs.~\ref{Re300_L21_middle} and \ref{Re500_L21_middle}). Here also, the dominating flow structure is well captured, mainly characterised by two jets appearing above the free inter-particle spaces of layer \#21 and leading to two vortices close to the walls. The recirculation zone with $z$-negative velocity direction in the center is also reproduced by the simulation, but is noticeably stronger in the experiment at $\mathrm{Re}_P = 300$. Overall, Fig.~\ref{Re300_L21_middle} shows the poorest comparison between simulation and experiment, which might indicate some unexpected perturbation coming from the outflow. The fluctuations discussed later in Sec.~\ref{Fluctuations} might also possibly explain at least part of this visible difference. At higher inflow velocity ($\mathrm{Re}_P = 500$, Fig.~\ref{Re500_L21_middle}) the high-speed jets dominate even more the resulting flow fields. The position of the left jet appears somewhat shifted toward the top-left corner in the simulation compared to PIV data, but the agreement is much better than at $\mathrm{Re}_P = 300$.

\begin{figure}[h!]
  \centering
   \hspace{-0.5cm}
 {\includegraphics[width=10cm,keepaspectratio]{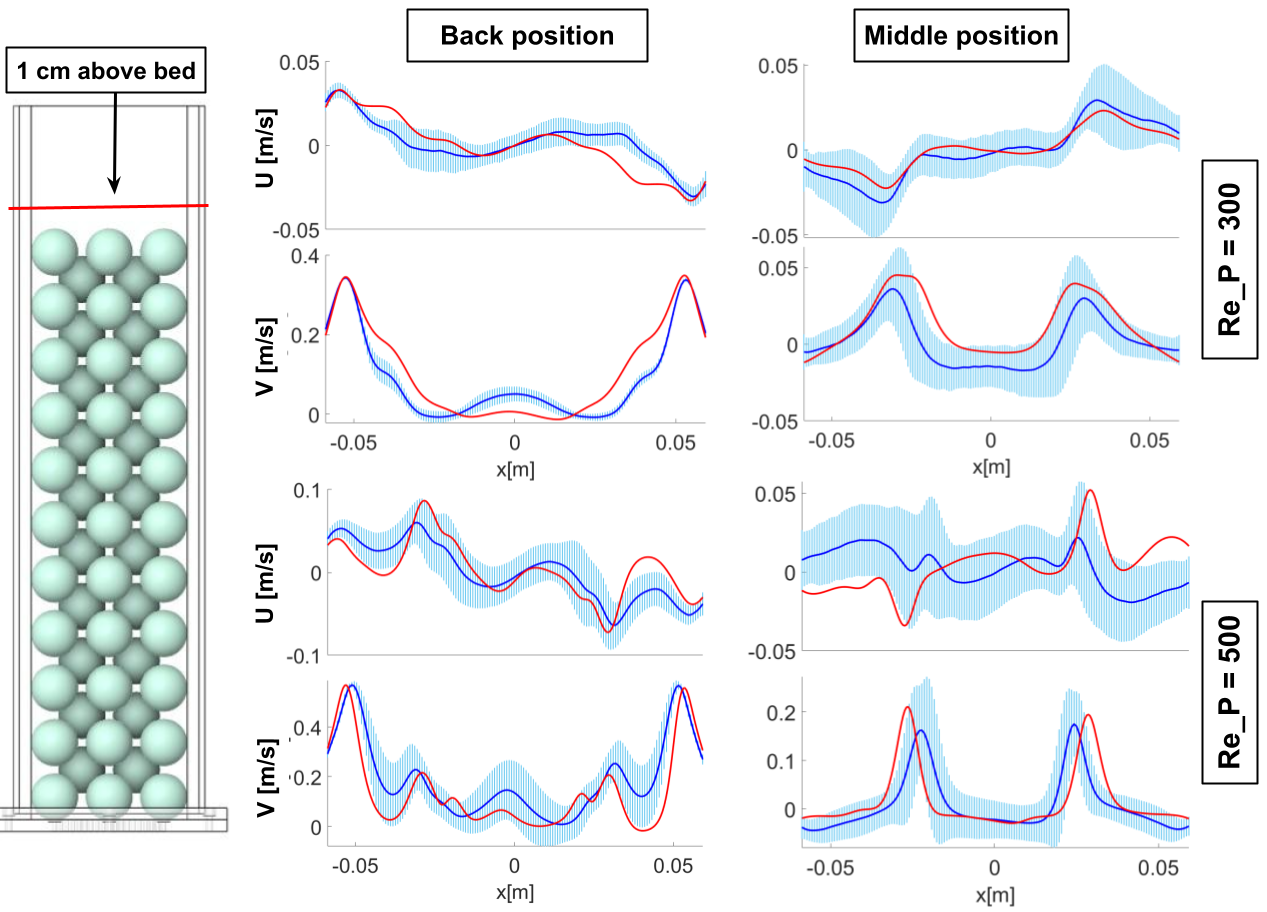}}
	\caption{Comparison of experimental (blue line) and numerical data (red line) for $\mathrm{Re}_P = 300$ (top half) and $\mathrm{Re}_P = 500$ (bottom half) 1~cm above the bed. For each case the horizontal velocity $u$ (top part of each subplot) and the vertical velocity $v$ (bottom part of each subplot) are shown, together with the $2\sigma$ confidence interval from PIV as a light-blue region. Note the different vertical scales.\label{Plots_h1}}
\end{figure}

To provide a more quantitative comparison, line plots for both, $\mathrm{Re}_P = 300$ and $\mathrm{Re}_P=500$, of the vertical ($v$) and horizontal ($u$) velocity components are now compared. These are the two velocity components measured by planar PIV within the light sheet. The comparisons are shown at both depth positions (back and middle)
1~cm above the particle bed (Fig.~\ref{Plots_h1}).
Additionally, the confidence intervals corresponding to twice the standard deviation ($2\sigma$) of the PIV data used for averaging, corresponding to a 95\% confidence level, are plotted. This allows a better quantification of the fluctuations observed experimentally from image to image.\\
Based on Fig.~\ref{Plots_h1}
it can be concluded that most of the flow structures are captured well.
Keeping in mind that the vertical scales (i.e., the peak velocity) differ from image to image, small dynamic ranges obviously magnify relative differences. In Fig.~\ref{Plots_h1}, the observed differences correspond mostly to a slight over- or underestimation of peak magnitude, and/or to a shift in positions, as already observed previously. This will be part of the discussion in the next section.

\subsection{Discussion of the comparisons and validation\label{sec:discus}}
Overall, a very good agreement has been obtained by comparing directly results of LBM simulations and experimental PIV measurements. Taking additionally into account the numerous previous verification and validation steps involving ALBORZ (discussed, among others, in \cite{TGV}) the numerical solver is, therefore, considered as being properly validated, and can now be used for further investigations, as described in the next section. The observed differences depend on Reynolds number, distance from the side walls, and position within or outside of the bed, justifying a more detailed discussion regarding the observed differences.

First, going from a particle Reynolds number of $\mathrm{Re}_P = 300$ to $\mathrm{Re}_P=500$, stronger discrepancies are observed between LBM and PIV in most cases. As will be discussed separately in Sec.~\ref{Fluctuations}, this is possibly due to low-frequency flow fluctuations connected to the onset of flow instabilities. Additionally, it must be kept in mind that different averaging times are used for PIV and LBM. The experimental data was recorded with a frequency of 2.5 Hz for a total of 300 images (for $\mathrm{Re}_P = 300$) or 1000 images (for $\mathrm{Re}_P = 500$), leading respectively to an averaging time of 120~s or even 400~s. On the other hand, the averaging time for the numerical simulation is only 32~s due to the very high computational costs coming with this large set-up. As a consequence, it is expected that the PIV results show the real averaged fields, while the rather short averaging time of LBM might suggest that the numerical results are not always fully converged to steady-state. This would also explain why some numerical results are still not symmetrical, while perfect symmetrical set-up and boundary conditions are considered in the simulation.

Flow fluctuations are probably also the reason why the comparisons are better within the bed than above it. Obviously, the flow direction is strongly constrained within the BCC packing, enforcing specific flow structures. Approaching the end of the particle bed (within the very last layers), and even more above it, this is not true anymore. The resulting flow becomes far more sensitive to intrinsic instabilities, as well as to possible perturbations induced at the outlet of the reactor/of the computational domain. The video showing the freeboard submitted as supplementary material to this article shows clearly the flow fluctuations appearing after leaving the packed bed, while the second video demonstrates that fluctuations in time are visually not apparent within the particle bed itself.

Finally, the better agreement observed farther from the side walls is expected to come from strong channelling effects, resulting first from the small number of spheres in cross-section direction (only $3\times 3$ in strong layers). Later studies will have to increase this number, and/or use 3D-printing techniques to include the missing half-spheres along the vertical walls. Additionally, all the geometrical uncertainties coming with the experimental set-up may play a role to magnify this problem. While the LBM simulations consider a perfect BCC packing with point-contact between the spheres/with the walls, the reactor used for PIV measurements involves unavoidable uncertainties, in particular: manufacturing tolerance of the spheres, leading to very small variations in diameter or imperfect sphericity, insufficient precision of construction or placement of the side walls, slightly varying wall thickness, walls being locally non-flat (bulging toward or away from the packing). As a consequence, small gaps could appear locally between packing and container walls, impacting noticeably channeling effects. Finally, the manual positioning of the spheres and the gluing process might induce slight deviations from an ideal BCC packing. The small glue bridges between the spheres are definitely larger than the contact points assumed in the simulations.

Keeping all those points in mind, the overall agreement observed between LBM simulations and PIV measurements is deemed very good, successfully terminating the validation step regarding lattice Boltzmann simulations of the flow through packed beds. The rest of this article deals with a deeper flow analysis using only numerical simulations.

\section{Further numerical studies regarding gas flows in BCC packing}
\subsection{Effect of the number of particle layers}
This section discusses the impact of the number of layers on the resulting flow structure at the end of and above the bed. Preliminary experimental studies revealed that an odd number of layers around 20 is necessary to get results that do not change when adding further particle layers; 17 was a minimum, 21 layers have been finally used to have a safety margin.

To check that point by simulations, the flow conditions obtained with various bed heights are now assessed, comparing the results obtained with 15, 17, 19, and 21 layers. In a packed bed reactor, it is well-known from the literature that the bed height affects pressure drop, channeling effects, flow distribution, and residence time of the fluid. If the bed height is too small, the reactor may not provide sufficient residence time for the desired reaction to occur, while a bed height that is too large will lead to excessive pressure drop and possibly poor flow distribution. Therefore, determining the minimum number of layers required for the flow to become independent of the bed height is crucial in ensuring optimal reactor performance.

\begin{figure}[ht!]
  \centering
   \hspace{-0.5cm}
 {\includegraphics[width=8cm,keepaspectratio]
  {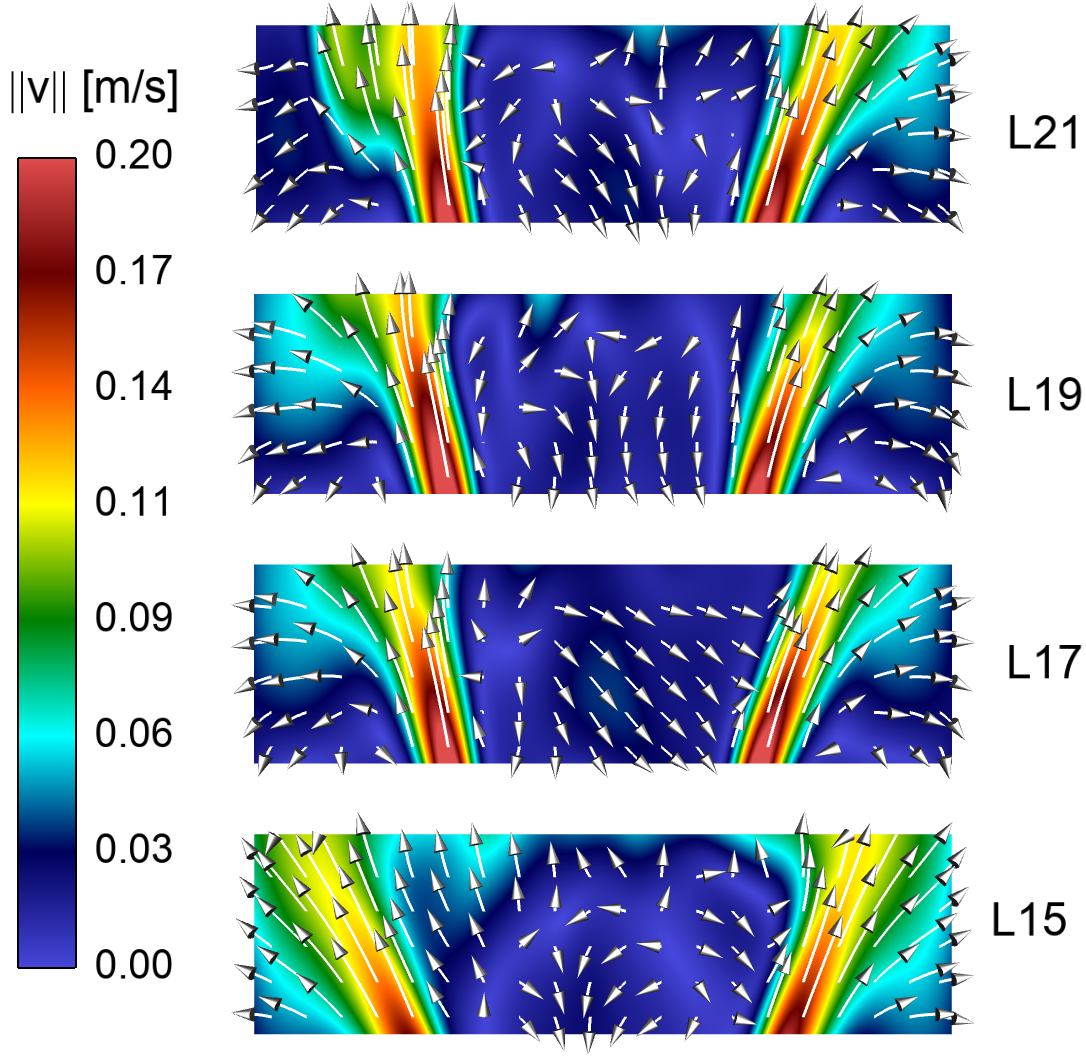}}
   \caption{Comparison of the average in-plane velocity magnitude in the central vertical plane ($y=0$) just above the BCC packing for $\mathrm{Re}_P=500$ with decreasing number of layers (from top to bottom): 21, 19, 17, 15.}
\label{layer_comparison}
\end{figure}

To this end, four distinct simulations were performed for $\mathrm{Re}_P=500$, using 15, 17, 19, or 21 layers, respectively. The highest bed with 21 layers corresponds to the configuration used up to now in this article. All simulations were conducted using identical boundary conditions, the same grid spacing and time-step, in order to ensure a fair comparison between the results. After careful analysis of the results, as confirmed visually by looking at Fig.~\ref{layer_comparison}, it becomes evident that the flow pattern does not change noticeably in the simulations when using an odd number of layers exceeding at least 17. While clear differences are observed in the flow above 15 or above 17 layers in the BCC packing, increasing further the total number of layers to 19 or 21 leads only to very small changes. This confirms the validity of our previous choice, using 21 layers for comparisons between PIV measurements and simulation data.

In Figure~\ref{layer_comparison} the average in-plane velocity magnitudes at $\mathrm{Re}_P = 500$ in the middle plane of the reactor containing 15, 17, 19 or 21 layers are presented. Note that, in this purely numerical study, all three velocity components are of course known and could be used to compute the velocity magnitude. In order to facilitate comparisons with the previous figures, a representation of the in-plane velocity magnitude (computed from the $x$ and $z$-components only) has been kept.

\begin{figure}[ht!]
	\centering
	\includegraphics[width=11cm,keepaspectratio]{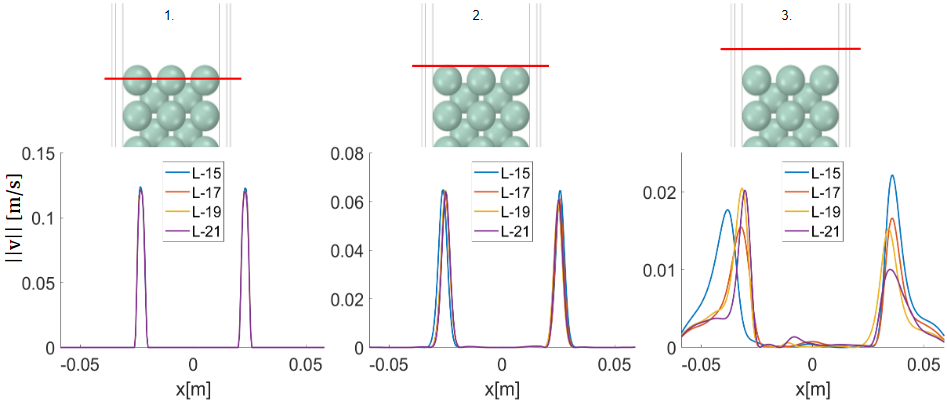}
	\caption{Comparison of in-plane velocity magnitudes in the center of the reactor ($y=0$) for $\mathrm{Re}_P = 500$ at 3 different heights when increasing the total number of layers from 15, 17, 19, up to 21 layers: (left) in the center plane of the last particle layer; (middle) just after leaving the particle bed; (right) 3 cm above the particle bed. Note the different vertical scales.}
	\label{Multilayer}
\end{figure}

To gain a more comprehensive understanding of these changes in flow velocity, Figure~\ref{Multilayer} provides a quantitative analysis of the in-plane velocity magnitude within and above the last particle layer, considering again 15, 17, 19, or 21 layers in total.
In Fig.~\ref{Multilayer}, the results have been plotted in the center of the reactor ($y=0$) at three different heights: 1) in the middle of the last particle layer, at the level of sphere center (left); 2) directly after exiting the particle bed (center); 3) 3~cm above particle bed (right figure). No noticeable differences are observed in the left figure, still located within the bed (middle of last particle layer). Just after leaving the particle bed (central part of Fig.~\ref{Multilayer}), the results obtained with 17, 19, and 21 layers are fully identical, while the blue curve (corresponding to 15 layers) already shows small but visible deviations. This confirms our previous statement -- at least 17 layers should be used to get exactly the same flow conditions up to leaving the particle bed. Finally, as the flow proceeds further in the free space above the bed, as shown in the right part of Fig.~\ref{Multilayer}, the differences amplify strongly. The results with a 15-layer bed look completely different there. Qualitatively, the velocity profiles with 17, 19, or 21 layers look the same, but visible differences start to appear as well, and the profiles are not fully symmetric any more. This will be the subject of the following last section.

\subsection{Velocity fluctuations} 
\label{Fluctuations}
In the course of this study, the experiments and a video of the numerical results (provided as supplementary material to this article) revealed that the gas flow, after exiting the packed bed, shows clear fluctuations in the freeboard region above the packing. Due to the relatively low particle Reynolds numbers involved in this study, the corresponding frequencies are quite low. They are, therefore, not easy to analyze accurately. Experimentally, PIV data was acquired at a low frequency, with no attempt to get full resolution in time. Regarding LBM simulations, the physical time that can be solved was up to now limited to 32 s due to the high cost of these computations.

In classical CFD, unsteady phenomena can be first identified by looking at the evolution of the residuals. Such residuals can of course also be computed in LBM simulations. Doing this, it was found that the residuals for velocity magnitude drop very sharply during the first 13~s (for $\mathrm{Re}_P = 300$) or 8~s (for $\mathrm{Re}_P=500$) of physical time before reaching a low-level, nearly constant value. This plateau, again, depends on the particle Reynolds number. While the normalized residuals oscillate slightly around $5\cdot10^{-6}$ at $\mathrm{Re}_P = 300$, they cannot drop below $4\cdot10^{-5}$ at $\mathrm{Re}_P = 500$, revealing larger variations of velocity with time at the higher particle Reynolds number. The analysis of flow fluctuations over time provides valuable insights regarding the dynamics of the flow, and is an important step for a better understanding of fluid flows in porous media~\citep{Boutt}.

In the previous comparisons and discussion, it was mentioned several times that low-frequency fluctuations have been observed in the numerical simulations, possibly explaining non-symmetrical flow features and differences from the experiments. For this reason, it was decided to pursue the LBM simulations for a noticeably longer physical time by accessing a High-Performance Computer. However, saving all the corresponding results on disk was not possible due to storage limitations. This is why only selected results can be discussed in what follows.

\begin{figure}[ht!]
\centering
\includegraphics[width=12cm,keepaspectratio]{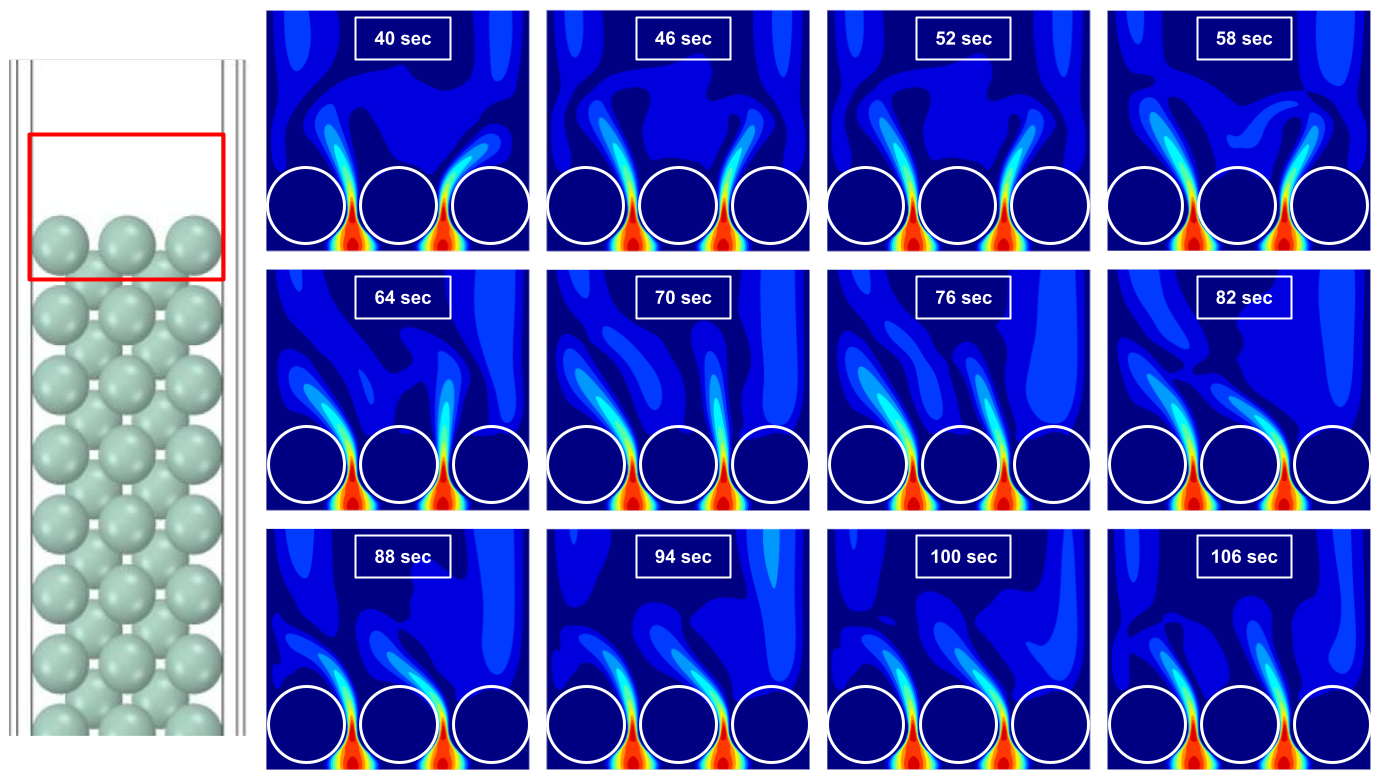}
\caption{Representation of the instantaneous in-plane velocity magnitude in the central symmetry plane ($y=0$) within the last particle layer and above the bed (see red rectangle on the left side) for a particle Reynolds number of 300. From top left to bottom right, 12 results are shown separated by 6~s of physical time each, leading to a total duration of 66~s.}\label{fluctuation}
\end{figure}

As a revealing example of the observed, low-frequency fluctuations, Figure~\ref{fluctuation} provides a visual representation of how the changes in flow structures take place with time. This figure shows the instantaneous in-plane velocity magnitude in the center plane of the reactor ($y=0$) within and above the last particle layer (layer \#21), in the region bounded by the red rectangle plotted on the left side. Results separated by 6~s of physical time during somewhat more than one minute in total (from 40 to 106~s) are shown from top left to bottom right. A very clear jet flapping phenomenon is observed here for $\mathrm{Re}_P = 300$. Interestingly, jet flapping is not symmetric, the right jet changing completely horizontal direction, while the left jet keeps its dominating orientation towards the wall in the simulated time windows of 66 seconds. It is expected that, if it would be possible to pursue this very costly simulation even further, symmetry would re-establish later on with a left jet switching as well its horizontal direction. Similar low-frequency fluctuations have been observed also for the higher particle Reynolds number.

Analyzing individual images, it appears possible that these flow movements are the results of weak vortex shedding in the wake of the top-layer particles. An oscillating flow pattern is created as a result of such vortices being shed alternatively on both sides of the obstruction. Due to the interactions between the shedding processes behind the different particles and with the side walls symmetry is broken, inducing complex flapping movements. The exact geometry of packing and reactor, fluid characteristics, flow rate and Reynolds number will impact the flapping behavior~\citep{Shi}. For all applications where the behavior of the flow after leaving the packed bed is important for the final process outcome, a better understanding of these low-frequency fluctuations -- or possibly finding a way how they could be suppressed or controlled -- would be important. This will be the subject of future work.

\begin{figure}[h!]
	\centering
	\includegraphics[width=8cm,keepaspectratio]{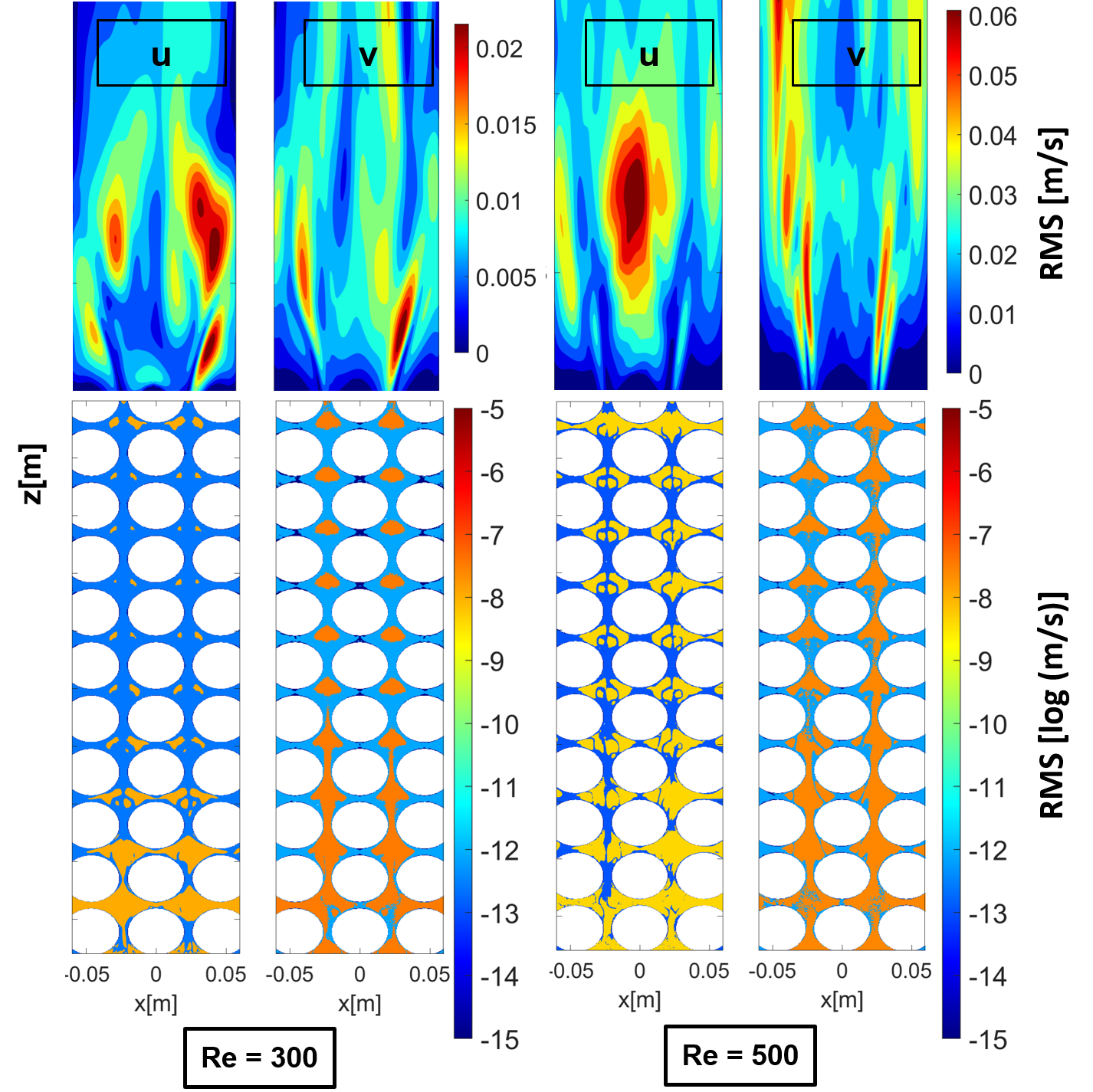}
	\caption{Root Mean Square (RMS) of velocity components $u$ (horizontal direction, left part of each subfigure) and $v$ (vertical direction, right part of each subfigure) for $\mathrm{Re}_P=300$ (left) or $\mathrm{Re}_P=500$ (right). The values inside the packed bed are represented with a logarithmic color scale in order to improve readability, while the much larger RMS values above the bed are shown with a linear scale, but with different scales for the two different Reynolds numbers.}
\label{rms}
\end{figure}

Root Mean Square (RMS) is a widely used statistical measure that quantifies the fluctuations in a signal. Here, RMS is used to analyze the magnitude of velocity fluctuations, both within the BCC packing and in the freeboard above it. Low RMS values indicate negligible fluctuations, while high values will be associated with noticeable unsteady features \citep{Kawamura}. The resulting values, presented in Fig.~\ref{rms}, have been computed separately for the horizontal ($u$) and vertical ($v$) in-plane velocity components. Again, the two particle Reynolds numbers $\mathrm{Re}_P = 300$ and $\mathrm{Re}_P=500$ have been considered. The results reveal that the flow above the packed bed, with RMS values plotted using a linear color scale, shows fluctuations orders of magnitude larger than within the BCC packing, in which a logarithmic scale must be used to be readable. This shows that, within the packed bed, fluctuations are extremely small except in the last particle layer. They do not increase much when going from $\mathrm{Re}_P = 300$ to $\mathrm{Re}_P=500$, but the regions showing noticeable fluctuations grow in size.

The situation at the exit of and above the bed is completely different, with fluctuations showing amplitudes exceeding locally 6~cm/s for $\mathrm{Re}_P=500$. Looking at the different color scales, it can be seen that the peak RMS values found above the bed are roughly three times higher for $\mathrm{Re}_P = 500$ than those for $\mathrm{Re}_P = 300$. This indicates that velocity variations with time become considerable after exiting the particle bed. A more detailed analysis of these velocity fluctuations by Proper Orthogonal Decomposition \citep{PODK,PODJ,mPOD} would be very interesting; however, POD necessitates acquisition (by PIV) and storage (for LBM simulations) of more snapshots over longer times.

\section{Conclusions}
The focus of this study is to develop and validate a numerical model that can accurately describe the behavior of gas flows through a packed bed reactor. To achieve this, an existing hybrid solver combining the lattice Boltzmann formulation with finite differences is improved further to examine the flow behavior of a model packed bed, particularly above and in-between the last layers of the BCC packing. For validation purposes, the results obtained from the numerical model have been directly compared with experimental data obtained by Particle Image Velocimetry for exactly the same configuration.\\
Detailed comparisons within the last 4 layers of the packed bed reveal very good agreement with the measurement data. The remaining, small discrepancies can be mostly attributed to minute differences regarding packing arrangement and reactor geometry in the real set-up, compared to the ideal geometry considered in the simulations, as well as to low-frequency fluctuations in time. Then, the validated numerical model has been used to investigate the effect of the number of particle layers on the flow structure at the outlet of the bed. This analysis revealed that a minimum number of 17 layers must be used to get always the same flow structures at the exit of the BCC packing, confirming preliminary experimental observations.\\
Interestingly, the study also revealed considerable velocity fluctuations in the freeboard just above the bed, even at low particle Reynolds numbers. At the same time, the magnitude of the velocity fluctuations within the particle bed is extremely low, except within the very last layer. Overall, these findings indicate that the lattice Boltzmann simulation is an effective tool for modeling and analyzing the fluid dynamics of such systems. To summarize:
\begin{itemize}
    \item A combined experimental (PIV) and numerical (LBM) study of the same configuration is a particularly attractive solution for cross-validation and to maximize our understanding of the main flow features in packed beds.
    \item The lattice Boltzmann solver developed in this study is suitable for getting accurate predictions of the resulting flow.
    \item A minimum of 17 particle layers must be used to ensure that the flow structures obtained at the outlet of the particle bed do not change any more.
    \item Even at such low particle Reynolds numbers as 300, very noticeable velocity fluctuations are observed at the outlet of and above the particle bed, associated to low-frequency oscillations of the flow structure.
\end{itemize}
The results of this study have the potential to enhance the understanding of gas dynamics in packed bed reactors. A future perspective would be to extend the methodology by taking into account heat transfer around non-spherical particles~\citep{Namdar23} and chemical reactions~\citep{Alireact}, with the ultimate objective of optimizing corresponding high-temperature conversion processes in packed beds.
\section*{Ethics declarations}
\newcommand{\circbullet}{\textbullet\hspace{1mm}}

\begin{itemize}
 \item[\circbullet] \textbf{Funding: }
This work has been funded by the Deutsche Forschungsgemeinschaft (DFG, German Research Foundation) within CRC/TRR 287 “BULK-REACTION” under number 422037413.\\

 \item[\circbullet] \textbf{Conflict of interest/Competing interests: }
The authors declare that they have no conflict of interest.\\

 \item[\circbullet] \textbf{Ethics approval: } Not applicable. The research did not involve any human or animal participants.\\

 \item[\circbullet] \textbf{Informed consent: } Not applicable. The research did not involve any human participants.\\

 \item[\circbullet] \textbf{Authors' contributions: }
Tanya Neeraj is in charge of all numerical simulations and writing of the article. Chirstin Velten is in charge of all experimental measurements and writing of the article. Gabor Janiga provided support for post-processing and image analysis. Katharina Zähringer provided scientific supervision of experimental work. Reza Namdar provided support for the article correction. Fathollah Varnik provided support for the article correction, project management. Dominique Thévenin provided support for the article correction, project management. Seyed Ali Hosseini provided scientific supervision of numerical work.
\end{itemize}

\section*{Acknowledgments}
This work has been funded by the Deutsche Forschungsgemeinschaft (DFG, German Research Foundation) within CRC/TRR 287 “BULK-REACTION” under number 422037413. Regarding final simulations on High-Performance Computers, the support of the Leibniz Supercomputing Center in Munich for accessing SuperMUC-NG is gratefully acknowledged.
\bibliographystyle{spbasic}
\bibliography{references.bib}
\end{document}